\documentclass[pra,aps,twocolumn,superscriptaddress,showpacs,tightenlines,amsmath]{revtex4}
\usepackage{epsfig}
\usepackage{amsmath,amssymb}

\begin{document}

\title{Algebraic and group structure for bipartite three dimensional anisotropic Ising model on a non-local basis}

\author{Francisco Delgado}
\email{fdelgado@itesm.mx}
\affiliation{Escuela Nacional de Posgrado en Ciencias e Ingenier\'ia, Tecnol\'ogico de Monterrey, M\'exico.}
\affiliation{Departamento de F\'isica y Matem\'aticas, Tecnol\'ogico de Monterrey, Campus Estado de M\'exico, Atizap\'an, Estado de M\'exico, CP. 52926, M\'exico.}

\date{\today}

\begin{abstract}

Entanglement is considered as a basic physical resource for modern quantum applications in Quantum Information and Quantum Computation theories. Interactions able to generate and sustain entanglement are subject to deep research in order to have understanding and control on it, based on specific physical systems. Atoms, ions or quantum dots are considered a key piece in quantum applications because is a basic piece of developments towards a scalable spin-based quantum computer through universal and basic quantum operations. Ising model is a type of interaction which generates and modifies entanglement properties of quantum systems based on matter. In this work, a general anisotropic three dimensional Ising model including an inhomogeneous magnetic field is analyzed to obtain their evolution and then, their algebraic properties which are controlled through a set of physical parameters. Evolution denote remarkable group properties when is analyzed in a non local basis, in particular those related with entanglement. These properties give a fruitful arena for further quantum applications and their control.


\pacs{03.67.-a; 03.67.Bg; 03.65.Ud; 03.65.Ge; 03.65.Fd; 03.65.Aa; 02.20.Uw}

\end{abstract} 

\maketitle

\section{Introduction}

Quantum entanglement is one of the most interesting properties of Quantum Mechanics which was noted since early times of theory \cite{vneumann1, schrod1, schrod2, einstein1, schrod3}. Nowadays, this property has been exploited by quantum applications as central aspect to improve information processing in terms of capacity and speed \cite{jozsa1, jozsa2, bennet1}. Thus, Quantum Information studies entanglement as an important aspect to codify and manage information in several quantum applications developed since seminal proposals in Quantum Computation \cite{feyn1, deutsch1, steane1}, Quantum Cryptography \cite{bennet2, ekert1} and discoveries about superdense coding \cite{bennet3} and teleportation \cite{bennet4}. A complete entanglement map of road will not be constructed until its quantification and behavior could be understood since a general mathematical theory and a deep knowledge about 
quantum interactions which generates it. It last means, Hamiltonian models which are able to generate entanglement, which are actually studied in order to understand how this quantum feature is generated on several physical systems. For magnetic systems, Ising model \cite{ising1,brush1} in statistical physics and Heisenberg model \cite{baxter1} in quantum mechanics are Hamiltonian models derived from interaction between spin systems when they include a magnetic field, it works as a driven element in Hamiltonian. Nielsen \cite{nielsen2} was the first reporting studies of entanglement in magnetic systems based on a two spin systems using that model including an external magnetic field. 

Magnetic driven Ising interaction is well known by developing an evolution depending on local parameters. Still its simplicity, for only two particles it exhibit four energy levels introducing a non periodical behavior in terms of Rabi frequencies phenomenon and their control \cite{kiri1,meng1}. Several simplified models has been analyzed in order to understand quantum behavior of these kind of systems when they approach to different concrete systems as quantum dots or electronic gases. Still, research around of control and entanglement in bipartite qubits \cite{terzis1} and lattices \cite{stelma1,novo1} is fundamental because these simple systems let the possibility to control  quantum states of a single or a couple of electron spins at time, standing at the heart of developments towards a scalable spin-based quantum computer.   

Control being depicted, in combination with controlled exchange between neighboring spins, would let obtain universal quantum operations \cite{recher1,saraga1,kopp1} in agreement with DiVincenzo criteria \cite{vincenzo1} in terms of reliability of state preparation and identification of well identified qubits. Thus, the aim of this paper is analyze algebraic properties of a bipartite system with a general three dimensional anisotropic Ising interaction including an inhomogeneous magnetic field strength in a  fixed direction. One of the central aspects is that analysis of dynamics is conducted on a non-local basis in terms of classical Bell states, which lets to discover outstanding algebraic aspects of this interaction around entanglement and a regular group structure, obtaining possible direct applications for quantum control and quantum computer processing.

\section{Anisotropic Ising model in three dimensions}

Different models of Ising interaction (XX, XY, XYZ depending on focus given by each author) has been considered in order to reproduce calculations related with bipartite and tripartite systems \cite{berman1,wang2,aless1}). Similar models which requires interaction with radiation are modeled in terms of Jaynes-Cummings and Jaynes-Cummings-Hubbard Hamiltonians \cite{meek1,rai1,ange1}. As example, in quantum control, different versions of Ising interaction have been considered in terms of homogeneity of magnetic field, dimensions and directions involved \cite{branczyk1,xi1,delgado1, delgado2}. Thus, restrictions in dimensions, number of particles and strength of external fields in these models are due for simplicity, geometry of lattices and other properties of physical systems involved \cite{wang2,kamta1,sun1,zhou1, gunlycke1}. 

In this work, we focus on the following Hamiltonian for the bipartite anisotropic Ising model \cite{ising1,nielsen2} including an inhomogeneous magnetic field restricted to the $h$-direction ($h=1,2,3$ corresponding with $x,y,z$ respectively):

\begin{eqnarray} \label{hamiltonian}
H_h&=&-\mathbf{\sigma_1 \cdot J \cdot \sigma_2}+\mathbf{B_1 \cdot \sigma_1}+\mathbf{B_2 \cdot \sigma_2} \nonumber \\
&=& -\sum_{k=1}^3 J_k {\sigma_1}_k {\sigma_2}_k+{B_1}_h {\sigma_1}_h+{B_2}_h {\sigma_2}_h\ 
\end{eqnarray}

\noindent which attempts to generalize most of several models considered in the cited works before. By diagonalizing and finding the corresponding eigenvalues, which are independent of $h=1,2,3$:

\begin{eqnarray} \label{eigenvalues}
{\mathcal{E}_h}^{(1)}=-J_h-{R_h}_+ &,& 
{\mathcal{E}_h}^{(2)}=-J_h+{R_h}_+ \\ \nonumber
{\mathcal{E}_h}^{(3)}= \quad J_h-{R_h}_- &,& 
{\mathcal{E}_h}^{(4)}= \quad J_h+{R_h}_- 
\end{eqnarray}

\noindent where ${R_h}_-$ and ${R_h}_+$ are defined as follows: if $h,i,j$ is a cyclic permutation of $1,2,3$ which will be simplified by using $\{ h \}$ symbol, being equivalent to the pair of indexes $i,j$  :

\begin{eqnarray} \label{erres}
{R_h}_\pm &=&\sqrt{{B_h^2}_\pm+{J_{i,j}^2}_\mp}=\sqrt{{B_h^2}_\pm+{J_{\{ h \}}^2}_\mp}
\\ \nonumber
{\rm with:} && J_{\{ h \}\pm} \equiv {J_{i,j}}_\pm=J_i \pm J_j \\ \nonumber 
&& B_{h \pm}=B_{1_h}\pm B_{2_h}
\end{eqnarray}

\subsection{Reduced notation and definitions}

One more suitable selection of reduced parameters will establish an appropriate notation which we will follow in the whole of work (in order to make finite some parameters and to reduce extent of some expressions):

\begin{eqnarray} \label{defs}
{b_h}_\pm=\frac{{B_h}_\pm}{{R_h}_\pm} &,& {j_h}_\pm=\frac{{J_{\{h\}}}_\mp}{{R_h}_\pm} \in [-1,1]
\end{eqnarray}

Note that subscripts $-,+$ are settled for these variables in relation with their internal operations in (\ref{defs}). It is known that when anisotropic Ising evolution matrix is expressed in the computational basis, it has in general a complex full form (with full 4 $\times$ 4 entries complex and different of zero), except if magnetic field is in $z$ direction \cite{delgado1,delgado2}, which denotes the privileged basis selected. For this reason, in the next section, the analysis will be based on Bell state basis to develop a different structure in it. This
selection suggests to change the notation by using $-,+$ for lower and upper scripts. Nevertheless, when these labels appear in mathematical expressions, it will be convenient recover them to express operations, so they will be assumed as $-1,+1$ respectively when they are forming part of algebraic expressions. Reader should be alert about it. In the same sense, capital scripts $A, B, ...$ will be reserved for $0, 1$ referred to the computational basis; greek scripts will reserved for $-1,+1$ or $-,+$; latin scripts $h,i,j,k,...$ will reserved spatial coordinates $x,y,z$ or $1,2,3$; and $a,b,c,...$ (between parenthesis) as scripts for energy levels $1,2,3,4$. $\cdot$ will be used sometimes to emphasize number multiplication between terms in scripts and avoid confusions. For example, in this notation we will write the standard Bell states as:

\begin{eqnarray} \label{bellnotation}
\left| \beta_{--} \right> \equiv \left| \beta_{00} \right> &,& 
\left| \beta_{-+} \right> \equiv \left| \beta_{01} \right> \\ \nonumber 
\left| \beta_{+-} \right> \equiv \left| \beta_{10} \right> &,&  
\left| \beta_{++} \right> \equiv \left| \beta_{11} \right>    
\end{eqnarray}

\subsection{Eigenvectors and Evolution operator}

In last terms, $\mathcal{E}_h^{(a)}: \mathcal{E}_h^{(1)},\mathcal{E}_h^{(2)},\mathcal{E}_h^{(3)},\mathcal{E}_h^{(4)}$ correspond with $E_{\mu \nu}: E_{--},E_{-+},E_{+-},E_{++}$ respectively and they could written in the current notation as:

\begin{eqnarray} \label{eigenvalues2}
{E_h}_{\mu \nu}&\equiv&{\mathcal{E}_h}^{(2+\mu+\frac{1+\nu}{2})}= \mu J_h+\nu {R_h}_{-\mu}  \\ \nonumber
& =&\mu J_h+\nu \sqrt{{B_h}^2_{-\mu}+{J^2_{\{h\}}}_{\mu}}
\end{eqnarray}

\noindent the corresponding eigenvectors for each direction $h$ are:

\begin{align} \label{eigenvectors}
\left| \phi_{\mu \nu}^{1} \right> =&\sum_ {\epsilon \in \{-,+\}} \frac {\delta_{+\epsilon}\nu (1+\mu \nu {j_1}_{-\mu})-\delta_{- \epsilon}\mu {b_1}_{-\mu} }{\sqrt{2}\sqrt{1+\nu\mu {j_1}_{-\mu}}} \left| \beta_{\mu \epsilon} \right> \nonumber \\ \nonumber 
\left| \phi_{\mu \nu}^{2} \right> =&\sum_{\epsilon \in \{-,+\}} \frac {\delta_{+\epsilon}i \nu (1+\mu \nu {j_2}_{-\mu})+\delta_{- \epsilon}\mu \epsilon {b_2}_{-\mu} }{\sqrt{2}\sqrt{1+\nu\mu {j_2}_{-\mu}}} \left| \beta_{\mu \cdot \epsilon \epsilon} \right> \\ \nonumber
\left| \phi_{\mu \nu}^{3} \right> =&\sum_{\epsilon \in \{-,+\}} \frac {(1+\nu {b_3}_{-\mu})+\nu \epsilon {j_3}_{-\mu} }{2\sqrt{1+\nu {b_3}_{-\mu}}} \left| \beta_{\epsilon \mu} \right> \nonumber \\
\end{align}

\noindent where $\delta_{\alpha \beta}$ is the custom Kronecker delta. 

An arbitrary bipartite state is written in computational basis or in Bell basis respectively as:

\begin{equation}\label{twoqubit}
\left| \psi \right>=\sum_{{A,B} \in \{ 0, 1 \}} \mathcal{A}_{A B} \left| A B \right> = \sum_{{\alpha, \beta} \in \{-, + \}} \mathcal{B}_{\alpha,\beta} \left| \beta_{\alpha \beta} \right> 
\end{equation}

\noindent then it is possible demonstrate by direct calculation that concurrence and Schmidt coefficients \cite{wott1}, in terms of coefficients $\mathcal{B}_{\alpha,\beta}$, are in the Bell basis:

\begin{eqnarray}\label{schmbell}
\mathcal{C}&=&\left| \sum_{{\alpha \beta} \in \{-, + \}} \beta \mathcal{B}^2_{\alpha, \alpha \cdot \beta} \right| \\ \nonumber \\
\lambda^2_{\left| \psi \right>}&=&\frac{1}{2} \left( 1 \pm \sqrt{1- \mathcal{C}^2}\right)
\end{eqnarray}

\noindent so, for eigenstates $\left| \phi_{\mu \nu}^i \right>$, coefficients are simply:

\begin{equation}
\lambda^2_{\left| \phi_{\mu \nu}^h \right>}=\frac{1}{2} ( 1 \pm \left| {b_h}_{-\mu} \right|)
\end{equation}

\noindent then, their entropy of entanglement becomes:

\begin{equation}
S_{\left| \phi_{\mu \nu}^h \right>}=1-\frac{1}{2} \sum_{\nu \in \{-, + \}} (1+\nu \left| {b_h}_{-\mu} \right|) \log_2 (1+\nu \left| {b_h}_{-\mu} \right|)
\end{equation}

\noindent which is maximal only if ${b_h}_\mu=0$ (symmetric or antisymmetric fields), in agreement with \cite{delgado1}.

\section{Form and structure of evolution operator}

Using last expressions for eigenvalues and eigenvectors, and introducing the following convenient definitions related with energy levels:

\begin{equation} \label{delta}
{\Delta_h}_\mu^\nu = \frac{t}{2} ({E_h}_{\mu +}+\nu {E_h}_{\mu -})=
\begin{cases}
\mu J_h t & \rm{if} \quad \nu=+ \\
{R_h}_{-\mu} t & \rm{if} \quad \nu=-
\end{cases}
\end{equation}

\noindent and:

\begin{eqnarray} \label{ed}
{e_h}_\alpha^\beta &=& \cos {\Delta_h}_\alpha^- + i \beta {j_h}_{-\alpha} \sin {\Delta_h}_\alpha^- \\ \nonumber
{d_h}_\alpha &=& {b_h}_{-\alpha} \sin {\Delta_h}_\alpha^-
\end{eqnarray}

\noindent then, if evolution operator is written in Bell basis as:

\begin{equation} \label{evop1}
U_{h}(t)=\sum_{\alpha,\beta,\gamma, \delta} {U_h}_{\alpha\beta,\gamma\delta} \left| \beta_{\alpha \beta} \right>\left< \beta_{\gamma \delta} \right|
\end{equation}

\noindent so, for those elements different from zero, its explicit form becomes :

\begin{align} \label{evop2}
&{U_{1}}_{\alpha \beta,\mu \nu}=\delta_{\alpha \mu} e^{i{\Delta_1}_\alpha^+}(\delta_{\beta \nu}{e_1}_\alpha^{\alpha \cdot \beta}-i \alpha (1-\delta_{\beta \nu}) {d_1}_\alpha) \\ \nonumber
&{U_{2}}_{\alpha \beta,\mu \cdot \alpha \nu \cdot \beta }=\delta_{\mu \nu} e^{i{\Delta_2}_{\alpha \cdot \beta}^+}(\delta_{+1 \mu}{e_2}_{\alpha \cdot \beta}^\alpha+\delta_{-1 \mu}\alpha {d_2}_{\alpha \cdot \beta}) \\ \nonumber
&{U_{3}}_{\alpha \beta,\mu \nu}=\delta_{\beta \nu} e^{i{\Delta_3}_\beta^+}(\delta_{\alpha \mu}{e_3}_\beta^{\alpha}+i (1-\delta_{\alpha \mu}) {d_3}_\beta) 
\end{align}

\noindent which gives close forms for evolution operators in each case when they are expressed in the non-local basis of Bell states. It express, in some sense, more explicitly the evolution of entanglement. Reader should note that scripts $-,+$ in variables defined in current section are related with energy labels more than internal operations (as in ${J_{\{h\}}}_\pm,{B_{\{h\}}}_\mp$) as before. Awareness about this detail in notation will avoid later misconceptions. \\
 
\subsection{Sector structure in the Evolution operator}

Last expressions could be appreciated better in matrix form:

\begin{widetext}
\begin{eqnarray} \label{mathamiltonian}
{U_{1}}(t)=& \left(
\begin{array}{c|c|c|c}
{e^{i {\Delta_1}_-^+}{e_1}_-^-}^* & i e^{i {\Delta_1}_-^+}{d_1}_-      & 0           & 0      \\
\hline
i e^{i {\Delta_1}_-^+}{d_1}_- & e^{i {\Delta_1}_-^+}{{e_1}_-^-}  & 0           & 0            \\
\hline
0         & 0          & {e^{i {\Delta_1}_+^+}{e_1}_+^+}^*  & -i e^{i {\Delta_1}_+^+}{d_1}_+  \\
\hline
0         & 0          & -i e^{i {\Delta_1}_+^+}{d_1}_+     & {e^{i {\Delta_1}_+^+}{e_1}_+^+} 
\end{array}
\right)  &\in \mathbb{S}_1 \\[3mm] \nonumber
{U_{2}}(t)=& \left(
\begin{array}{c|c|c|c}
e^{i {\Delta_2}_+^+}{{e_2}_+^+}^*    &   0   &   0   & - e^{i {\Delta_2}_+^+}{d_2}_+  \\
\hline
0  &  e^{i {\Delta_2}_-^+}{{e_2}_-^+}^* &  -e^{i {\Delta_2}_-^+}{{d_2}_-}  & 0        \\
\hline
0  &  e^{i {\Delta_2}_-^+}{{d_2}_-} &  e^{i {\Delta_2}_-^+}{{e_2}_-^+}  & 0           \\
\hline
e^{i {\Delta_2}_+^+}{d_2}_+    &   0   &   0   & e^{i {\Delta_2}_+^+}{{e_2}_+^+}  
\end{array} 
\right) &\in \mathbb{S}_2 \\[3mm] \nonumber
{U_{3}}(t)=& \left(
\begin{array}{c|c|c|c}
e^{i {\Delta_3}_-^+}{{e_3}_-^+}^* & 0 & i e^{i {\Delta_3}_-^+}{d_3}_-      & 0        \\
\hline
0  &  e^{i {\Delta_3}_+^+}{{e_3}_+^+}^* & 0 & i e^{i {\Delta_3}_+^+}{d_3}_+           \\
\hline
i e^{i {\Delta_3}_-^+}{d_3}_- & 0 &  e^{i {\Delta_3}_-^+}{{e_3}_-^+}      & 0         \\
\hline
0  &  i e^{i {\Delta_3}_+^+}{d_3}_+ & 0 & e^{i {\Delta_3}_+^+}{{e_3}_+^+} 
\end{array}
\right) &\in \mathbb{S}_3
\end{eqnarray}
\end{widetext}

$U_{h=1,2,3}(t)$ clearly have a $2 \times 2$ sector structure (sectors are not consecutive to form matrix blocks, instead, each non zero entry is a vertex of  square sectors embed in the whole matrix). Note that ${\mathbb I}_4$ is included by set simply $t=0$. By the properties of ${{e_h}_\alpha^\beta}$ and ${d_h}_\alpha$, sectors are unitary with $e^{2i {\Delta_h}_\alpha^+}$ as determinant. Because $U_h(t)$ is unitary, inverses are obtained just by take $U^\dagger_h(t)$ (nevertheless those structure, it is required prove if it can be obtained as $U_h(t')$ for some other physical parameters for the same system). In addition, as the sum of eigenvalues is zero, then $U(t) \in SU(4)$, which is an important aspect of our evolution operator.

Referring only to their structure, $4 \times 4$ special unitary matrices in $SU(4)$ formed by unitary $2 \times 2$ sectors in $U(2)$ as is depicted in (\ref{mathamiltonian}) (clearly with unitary and reciprocal determinants), they form groups. Thus, it is easy show that each one, $\mathbb{S}_1, \mathbb{S}_2, \mathbb{S}_3$, are subgroups of $SU(4)$ (identity, inverses and multiplication are included in each $\mathbb{S}_h$ and product of two elements in the set remains in it):

\begin{eqnarray}\label{subgroups}
\mathbb{S}_1 &=& \{ A \in SU(4) | A_{\alpha \beta, \gamma \delta}= \delta_{\alpha \gamma} u_{\alpha \beta, \gamma \delta}, \nonumber \\  
&& \quad (u_{\gamma \alpha, \gamma \beta})\vline_{\gamma=\pm} \in U(2)\} \\ \nonumber
\mathbb{S}_2 &=& \{ A \in SU(4) | A_{\alpha \beta, \gamma \delta}= \delta_{\alpha \cdot \gamma \beta \cdot \delta} u_{\alpha \beta, \gamma \delta} , \nonumber \\  
&& \quad (u_{\alpha \beta, \gamma \cdot \alpha \gamma \cdot \beta})\vline_{\gamma=\pm} \in U(2)\} \\ \nonumber
\mathbb{S}_3 &=& \{ A \in SU(4) | A_{\alpha \beta, \gamma \delta}= \delta_{\beta \delta} u_{\alpha \gamma, \beta \gamma} , \nonumber \\  
&& \quad (u_{\gamma \alpha, \gamma \beta})\vline_{\gamma=\pm} \in U(2)\} \\ \nonumber
\end{eqnarray}

\noindent where $\alpha, \beta, \gamma, \delta \in \{-, + \}$. In addition, we state the symbol ${\mathbb S}^*_h \subset {\mathbb S}_h$ to each set of matrices able to be generated by $U_h(t)$ in (\ref{mathamiltonian}). In them, the general structure for  their $2 \times 2$ sectors in $U(2)$ is:

\begin{equation}\label{sector}
{s_h}_{j} = {e^{i {\Delta_h}_\alpha^+} \left(
\begin{array}{cc}
{{e_h}_\alpha^\beta}^* & -q i^h {d_h}_\alpha   \\
q {i^*}^h {d_h}_\alpha & {{e_h}_\alpha^\beta}    
\end{array}
\right) \vline}_{\tiny \begin{aligned} \alpha &= (-1)^{h+j+1} \\ \beta &= (-1)^{j(h+l_j-k_j+1)} \\ q &= \beta (-1)^{h+1} \end{aligned}}
\end{equation}

\noindent where $h$ denotes the associated spatial coordinate of magnetic field, $j=1, 2$ is an ordering label for sector as it appears in the rows of the evolution matrix, corresponding with $k_j, l_j$, the labels for its rows in each matrix of (\ref{mathamiltonian}) (by example, $k_2=3, l_2=4$ are the labels for the second sector, $j=2$, in $U_{h=1}(t)$ it means ${s_2}_1$. Note particularly that determinant for each sector, $\det ({s_h}_{j})={e^{2i {\Delta_h}_\alpha^+}}$, are reciprocal because ${\Delta_h}^+_{-\alpha}=-{\Delta_h}^+_\alpha$. Note that ${s_h}_j \in U(2)$ as was previously stated, but not all elements of $U(2)$ is a ${s_h}_j$ for any parameters ${{e_h}_\alpha^\beta}, {d_h}_\alpha$ (because the form of $-q i^h {d_h}_\alpha, q {i^*}^h {d_h}_\alpha$ in entries $1,2$ and $2,1$ respectively, which not a general complex number because their arguments are limited to integer or half integer factors of $\pi$). This implies that ${s_h}_j$ is not necessarily a subgroup of $U(2)$, which open opportunities to extend their coverage in $U(2)$ with two o more pulses.

For a further discussion, it is notable to write the generic sector in exponential form in terms of Pauli matrices:

\begin{eqnarray}\label{shjexp}
{s_h}_{j} &=& {e^{i {\Delta_h}_\alpha^+}} e^{-i {\Delta_h}_\alpha^- {\bf n} \cdot \boldsymbol{\sigma}} \equiv {e^{i {\Delta_h}_\alpha^+}} {s_h}_{j,0} \nonumber \\
&=& {e^{i {\Delta_h}_\alpha^+}} \left( \cos {{\Delta_h}_\alpha^-} {\mathbb I}_2 - i \sin {{\Delta_h}_\alpha^-} {\bf n} \cdot {\boldsymbol \sigma} \right) \\
{\rm with:} &{\bf n}& = (q {{b_h}_{-\alpha}} \sin \frac{h \pi}{2}, q {{b_h}_{-\alpha}} \cos \frac{h \pi}{2}, \beta {{j_h}_{-\alpha}} ) \nonumber
\end{eqnarray}

\noindent where ${\boldsymbol \sigma}=(\sigma_1,\sigma_2,\sigma_3)$, $\mathbb{I}_n$ is the $n \times n$ identity matrix and ${s_h}_{j,0} \in SU(2)$ is the matrix sector with lacking of its numeric exponential factor, which is defined by further convenience. That structure shows that each kind of interaction (with external magnetic field in $h- \rm{coordinate}$) exclusively transforms Bell states in specific pairs as a $SU(2)$ operation plus a phase term in $U(1)$. Then, each sector for a given $h$ is responsible to combine linearly two Bell states $\left| \beta_{\mu \nu} \right>$ and $\left| \beta_{\mu' \nu'} \right>$ under a $U(1) \times SU(2)$ operation, with:

\begin{eqnarray}
(\mu' , \nu') &=& 
\left\{
\begin{array}{lc}
(\mu , \nu \otimes 1)  & , h=1 \\ 
(\mu \oplus 1, \nu \oplus 1) & , h=2 \\  
(\mu \oplus 1, \nu ) & , h=3  
\end{array} \right. 
\end{eqnarray}

\noindent where $\oplus$ is the sum module 2. Thus, it is possible associate, by eliminating the sector phase $e^{i {\Delta_h}_\alpha^+} \in U(1)$ and instantaneous exponential factor in component $\left| \beta_{\mu \nu} \right>$ (it is equivalent to use a certain rotating frame to whole system), a Bloch sphere between $\left| \beta_{\mu \nu} \right>$ and $\left| \beta_{\mu' \nu'} \right>$ in which, each part of $\left| \psi \right>$ in (\ref{twoqubit}) corresponding with each sector:

\begin{eqnarray}\label{psij}
\left| \psi_{j} \right> &=& \sum_{{\gamma,\delta} \in \{\mu \nu,\mu' \nu'\}} \mathcal{B}_{\gamma \delta} \left| \beta_{\gamma \delta} \right> \\
&\Rightarrow& \left| \psi \right> = \sum_{j = 1,2} \left| \psi_{j} \right> \nonumber
\end{eqnarray} 

\noindent evolves 'locally' driven by ${s_h}_{j}$ in sector $j$. Note that this evolution does not change probabilities between parts in each sector, but introduces relative phases because $e^{i {\Delta_h}_\alpha^+} \in U(1)$ and the complex mixing generated by ${s_h}_{j,0}$. Clearly ${s_h}_{j,0}$ are elements of a Lie group with parameters ${{\Delta_h}_\alpha^-}{\bf n}$, which will be important further.

\begin{figure}[pb] 
\centerline{\psfig{width=3.3in,file=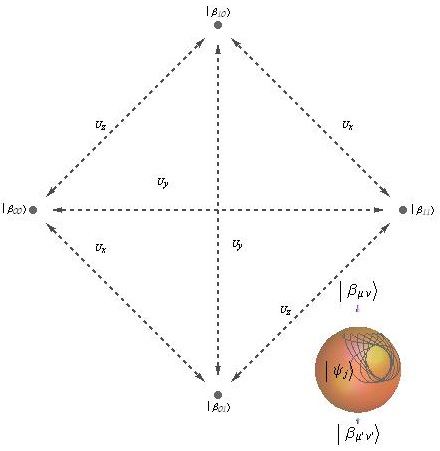}}
\vspace*{8pt}
\caption{Schematic representation of states related between each interaction in (\ref{mathamiltonian}). Each dotted line should be understood as linear combination of states on vertex. Each line represents an space equivalent to a Bloch sphere for related states $\left| \beta_{\mu \nu} \right>$ and $\left| \beta_{\mu' \nu'} \right>$. On this sphere, evolution of linear combination could be represented as lines depicted in the picture on the right driven by an operator in $SU(2)$.} \label{fig0}
\end{figure}

Figure \ref{fig0} shows Bell states related with several interactions $U_x(t), U_y(t), U_z(t)$ or $U_1(t), U_2(t), U_3(t)$ in (\ref{mathamiltonian}). Each dotted line is a linear combination of states in their vertex (\ref{psij}). Given a initial state of this type and depending on interaction being considered ($h=1, 2, 3$), evolution, for this part of quantum state, develops on a Bloch sphere as a trajectory, depicted in scheme on right-down, which shows its evolution trajectory for some pair $\left| \beta_{\mu \nu} \right>$ and $\left| \beta_{\mu' \nu'} \right>$ related by an specific interaction in their respective Bloch sphere.

By combining several adequate interactions, could be possible switch any Bell state into another (Figure 1) if sectors ${s_h}_{j}$ adopt the following combined forms for $t=T$ (referred in the following as diagonalization and antidiagonalization cases respectively):

\begin{eqnarray} 
{s_h}_{j} &=& \pm \mathbb{I}_2 \label{evloop} \\ \nonumber \\
{s_h}_{j}=\pm {\sigma_1} \quad &{\rm or:}& \quad {s_h}_{j}=\pm i \sigma_2 \label{exch}
\end{eqnarray} 

Thus, by combining these two types of sector forms, we can achieve evolution loops \cite{mielnik1, fernandez1, delgado3, delgado4} and exchange operations \cite{delgado2} in $\mathcal{H}^{\otimes 2}$ for Bell states, which will let obtain several control effects. Note that last expressions give only the matrix form for one sector ${s_h}_{j}$, avoiding a confusion between operators in computational basis and Bell basis by the use of Pauli matrices $\sigma_1,\sigma_2$ in last expression, which are used only to set a desired form in that matrix sector. Clearly, much more equivalent cases could be considered with arbitrary phases instead of only $\pm 1$ phases, but we will restrict our analysis to these cases.

\subsection{Evolution of Bell states entanglement}

In spite of (\ref{schmbell}) and (\ref{mathamiltonian}), concurrence of states evolved from Bell states could be easily obtained. Short calculations show that concurrence ${\mathcal C}^h_{\mu \nu}$ for evolution $U_h(t)$ for Bell state $\left|\beta_{\mu \nu} \right>$ is easily expressed as:

\begin{eqnarray}\label{bellentanglement}
{\mathcal C}^h_{\mu \nu} &=& 1-4 {j_h}_{-f^h_{\mu \nu}}{b_h}_{-f^h_{\mu \nu}} \sin^4 {\Delta_h^-}_{f^h_{\mu \nu}} \\
&{\rm with:}& f^h_{\mu \nu} = 
\left\{
\begin{array}{cc}
\mu & , h=1 \\ 
\mu \nu & , h=2 \\  
\nu & , h=3  
\end{array} \right. \nonumber
\end{eqnarray}

\noindent showing that it depends only on one Rabi frequency at time in a very simply way. This expression is consistent with isotropic case reported in \cite{delgado1,delgado2}. Note that (\ref{bellentanglement}) reproduces those results because this expression does not imply that states comeback each period to the original Bell state, only to the same entanglement value. Clearly, Bell states could become separable intermediately if ${j_h}^2_{f^h_{\mu \nu}} {b_h}^2_{f^h_{\mu \nu}}=1/4$, which is possible if ${\mathcal C}^h_{\mu \nu}$ reaches its maximum value when $|{B_h}_{-f^h_{\mu \nu}}|=|{J_{\{h\}}}_{f^h_{\mu \nu}}|$. In isotropic cases, some Bell states are invariant when magnetic field is symmetric or antisymmetric, as was shown in \cite{delgado2}, which in general does not happen here. All this scenario contrasts
with evolution of some separable states, by example those of computational basis $\left| ij \right>$, whose entanglement expressions depends normally on all or several Rabi frequencies involved, generally resulting in a non periodical behavior. Thus, in some sense, entanglement instead of separability appears as a natural feature of Ising model inclusive with external magnetic fields.

\subsection{Equivalence under rotations}
Of course, last evolution operators are related via an homogeneous bipartite rotation in terms of Euler angles \cite{sak1,morr1} on Hilbert space $\mathcal{H}$ and on Fock space $\mathcal{H}^{\otimes 2}$:

\begin{equation}
{R_{1,2}}(\alpha, \beta, \gamma)={R_1}(\alpha, \beta, \gamma) \otimes {R_2}(\alpha, \beta, \gamma)
\end{equation}

\noindent where:

\begin{eqnarray}\label{rotation}
R_i(\alpha, \beta, \gamma) &=&  \cos\frac{\beta}{2} \cos\frac{\alpha+\gamma}{2} \mathbb{I}_2 \\ \nonumber
&&- i \sin\frac{\beta}{2} \sin\frac{\alpha-\gamma}{2} \mathbf {\sigma}_1 \\ \nonumber
&&- i \sin\frac{\beta}{2} \cos\frac{\alpha-\gamma}{2} \mathbf {\sigma}_2 \\ \nonumber
&&- i \cos\frac{\beta}{2} \sin\frac{\alpha+\gamma}{2} \mathbf {\sigma}_3 \\ \nonumber
&=& \left(
\begin{array}{cccc}
e^{-i \frac{\alpha+\gamma}{2}}\cos \frac{\beta}{2} & -e^{i \frac{\alpha-\gamma}{2}}\sin \frac{\beta}{2}  \\
e^{-i \frac{\alpha-\gamma}{2}}\sin \frac{\beta}{2} & \quad e^{i \frac{\alpha+\gamma}{2}}\cos \frac{\beta}{2}
\end{array}
\right)
\end{eqnarray}

\noindent expressed in the computational basis. As is expected, different Ising models with magnetic fields in cartesian
directions transform between them. Specifically:

\begin{eqnarray}
{U_{1}}(t)&=&{R_{1,2}}(\frac{\pi}{2},\frac{\pi}{2},0){U_{3}}(t){R_{1,2}}^\dagger(\frac{\pi}{2},\frac{\pi}{2},0)   \\ 
{U_{2}}(t)&=&{R_{1,2}}(-\pi,\frac{\pi}{2},\frac{\pi}{2}){U_{3}}(t){R_{1,2}}^\dagger(-\pi,\frac{\pi}{2},\frac{\pi}{2}) \nonumber
\end{eqnarray}

An aspect to remark is that sectors (\ref{sector}) in Ising evolution matrix could reproduce the form of (\ref{rotation}). Nevertheless that single isolated qubit rotations by magnetic fields are well known, here this is not a trivial aspect
because we are using a representation in Bell basis instead of computational basis as in last expression. It means that, under certain conditions, Bell states rotate as a whole. This is not necessarily a physical rotation, but Bell states, by pairs corresponding to rows where that sector is located, are transforming between them under these specific driven Ising interactions as a rotation in a Bloch sphere with main states as those Bell states instead of classical $\left| 0 \right>$ and $\left| 1 \right>$.

\section{Group structure of evolution operators}

Previously has been stated that evolution operators $U_h(t)$ are part of subgroup ${\mathbb S}_h$ in $SU(4)$ defined by their form in (\ref{subgroups}). But there are a inner structure which can be found in terms of group properties, which are not only related with the form of these operators but instead with their quantum structure related with Ising Hamiltonian (\ref{hamiltonian}). In this section, we analyze specific structure and restrictions in (\ref{mathamiltonian}) to ${\mathbb S}^*_h \subset {\mathbb S}_h \subset SU(4)$ becomes a subgroup, together with traditional operator or matrix product in terms of their physical properties. It means, the physical prescriptions on $t, {j_h}_{\pm \alpha}, {b_h}_{\pm \alpha}$ parameters for each sector with which $U_h(t)$ fulfills a group structure:

\begin{eqnarray}
\begin{tabular}{l l l}\label{group}
{\rm Closure:} & &  \label{closure} \nonumber \\
$ U_h(t')U_h(t)  = U_h(t'') \in {\mathbb S}^*_h$ &&  \\
$\quad \quad \forall \quad U_h(t),U_h(t') \in {\mathbb S}^*_h$ && \\ \nonumber \\
{\rm Associativity:} \label{asoc} & & \nonumber \\
$U_h(t'') \left( U_h(t')U_h(t) \right) = \left( U_h(t'')U_h(t') \right) U_h(t)$ &&  \\
$\quad \quad \forall \quad U_h(t),U_h(t'),U_h(t'') \in {\mathbb S}^*_h$ && \\ \nonumber \\
{\rm Identity:} & & \label{identity2} \nonumber \\
${\mathbb I}_4 \in {\mathbb S}^*_h$ && \\  \\
{\rm Inverse:} \label{inverse} & & \nonumber \\
$U^{-1}_h(t) =U_h(t') \in {\mathbb S}^*_h$ &&  \\
$\quad \quad \forall \quad U_h(t) \in {\mathbb S}^*_h$ &&
\end{tabular} 
\end{eqnarray} 

\noindent where $U_h(t)$ is understood to have the structure in (\ref{mathamiltonian}). Clearly associativity and existence of identity (${\Delta_h}_{\pm \alpha}^+=2m \pi, {\Delta_h}_{\pm \alpha}^-=2n \pi, m,n \in {\mathbb Z}$) are fulfilled because covering $SU(4)$ structure. Thus, only the product closure and existence of inverse should be analyzed. Because ${\mathbb S}^*_h \subset {\mathbb S}_h$, sector structure is accomplished and analysis of previous conditions can be almost restricted to sectors ${s_h}_j$ (note only that exponential factor in each sector is the inverse of its respective factor in other sector, so both should be compatible in addition).  

\subsection{Inverse} 

In spite of sector properties of matrices in ${\mathbb S}^*_h$, inverse of $U_h(t)$ reduces to obtain inverse of each sector (caring compatibility around of their exponential factors). Because generic sector (\ref{sector}) is unitary, its inverse is:

\begin{equation}\label{inversesector}
{s_h}_{j}^{-1} = {e^{-i {\Delta_h}_\alpha^+} \left(
\begin{array}{cc}
{{e_h}_\alpha^\beta} & q i^h {d_h}_\alpha   \\
-q {i^*}^h {d_h}_\alpha & {{e_h}_\alpha^\beta}^*    
\end{array}
\right) }
\end{equation}

\noindent thus, conditions for a generic sector (\ref{sector}) mimicking last expression can be obtained by comparing entries $1,1$ with $2,2$ and $1,2$ with $2,1$ in ${s'_h}_{j}={s_h}_{j}^{-1}$. This comparison shows that there are two possible restrictions to make compatible those four equations:

\begin{eqnarray}\label{inversesectorcond1}
{\Delta'_h}_\alpha^+ + {\Delta_h}_\alpha^+ &=& p_\alpha \pi, p_\alpha \in {\mathbb Z} \label{inversesectorcond1A} \\
&{\rm or:}& \nonumber \\
{e_\alpha^\beta} {{e'_\alpha}^\beta} &=& {{e_\alpha^\beta}^*} {{{e'_\alpha}^\beta}^*} \label{inversesectorcond1B}
\end{eqnarray}

\noindent with which, only two additional equations remains, by example for entries $1,1$ and $1,2$:

\begin{eqnarray}\label{inversesectorcond2}
1,1: && {e^{i ({\Delta_h}_\alpha^+ + {\Delta'_h}_\alpha^+)}} {{e'_h}_\alpha^\beta}^* = {{e_h}_\alpha^\beta} \label{inversesectorcond2a} \nonumber \\
1,2: && -{e^{i ({\Delta_h}_\alpha^+ + {\Delta'_h}_\alpha^+)}} {d'_h}_\alpha = {d_h}_\alpha \label{inversesectorcond2b} \nonumber \\
\end{eqnarray}

Equation (\ref{inversesectorcond1A}) automatically fulfills the compatibility between sectors because exponential factor becomes real. Combining this condition (\ref{inversesectorcond2}), we get several cases. The most general is obtained noting that in spite of (\ref{inversesectorcond2b}), ${b'_h}_{-\alpha}={P_\alpha} {b_h}_{-\alpha}$. After, to fulfill (\ref{inversesectorcond2a}), it is required that ${\Delta'_h}_\alpha^- + S_\alpha {\Delta_h}_\alpha^- = n_\alpha \pi$ which implies ${j'_h}_{-\alpha}={S_\alpha} {j_h}_{-\alpha}$, $P_\alpha S_\alpha =1$ and $p_\alpha=2n_\alpha+m_\alpha$. With these conditions, ${s'_h}_\alpha$, with form (\ref{sector}), converts into (\ref{inversesector}). While, a brief analysis of equation (\ref{inversesectorcond1B}) into (\ref{inversesectorcond2}), shows that it reduces to a special case of last solution. Thus, the general prescriptions to get the inverse effect of an Ising interaction with a similar interaction but changing physical parameters are:

\begin{eqnarray}\label{inverseprescriptions}
{j'_h}_{-\alpha} &=& {S_\alpha} {j_h}_{-\alpha} \nonumber \\
{b'_h}_{-\alpha} &=& {S_\alpha} {b_h}_{-\alpha} \nonumber \\
{\Delta'_h}_\alpha^+ + {\Delta_h}_\alpha^+ &=& (2n_\alpha+m_\alpha) \pi \nonumber \\
{\Delta'_h}_\alpha^- + S_\alpha {\Delta_h}_\alpha^-  &=&  m_\alpha \pi  \\ \nonumber \\
{\rm with:} && S_\alpha  = \pm 1, n_\alpha, m_\alpha \in {\mathbb Z} \nonumber
\end{eqnarray}

Note that last prescriptions are compatible with prescriptions of evolution loops in two pulses. Still, there is a pair of particular solutions. The first case, where $P_\alpha S_\alpha=\pm 1$ corresponds to matrices with ${\Delta'_h}_\alpha^+ + {\Delta_h}_\alpha^+ = m_\alpha \pi, {\Delta_h}_\alpha^-=r_\alpha \pi, {\Delta'_h}_\alpha^-=(2n_\alpha + r_\alpha +m_\alpha) \pi$ with $n_\alpha, m_\alpha, r_\alpha \in {\mathbb Z}$ which is the form of sectors in diagonal form ${\mathcal D}_h^{{\Delta_h}_\alpha^+}$, whose inverse is simply ${\mathcal D}_h^{-{\Delta_h}_\alpha^+}$:

\begin{eqnarray} \label{graldiag2}
{U_{1}}(T)&\equiv& {\mathcal D}_1^\phi = \left(
\begin{array}{c c c c}
S_{1,1}e^{i \phi} & 0 & 0 & 0      \\
0 & S_{1,1}e^{i \phi} & 0 & 0            \\
0 & 0 & S_{1,2}e^{-i \phi} & 0  \\
0 & 0 & 0 & S_{1,2}e^{-i \phi} 
\end{array}
\right) \nonumber \\
{U_{2}}(T)&\equiv& {\mathcal D}_2^\phi = \left(
\begin{array}{c c c c}
S_{2,1}e^{i \phi} & 0 & 0 & 0      \\
0 & S_{2,2}e^{-i \phi} & 0 & 0            \\
0 & 0 & S_{2,2}e^{-i \phi} & 0  \\
0 & 0 & 0 & S_{2,1}e^{i \phi} 
\end{array}
\right) \nonumber \\
{U_{3}}(T)&\equiv& {\mathcal D}_3^\phi = \left(
\begin{array}{c c c c}
S_{3,1}e^{i \phi} & 0 & 0 & 0      \\
0 & S_{3,2}e^{-i \phi} & 0 & 0            \\
0 & 0 & S_{3,1}e^{i \phi} & 0  \\
0 & 0 & 0 & S_{3,2}e^{-i \phi} 
\end{array}
\right) \nonumber \\ \nonumber \\
\end{eqnarray}

\noindent where $S_{h,j}$ are $\pm1$ independently. With exception of freedom in selection of sign in $P_\alpha S\alpha$, these prescriptions agree with (\ref{inverseprescriptions}).

An possible additional case, which appears when equations (\ref{inversesectorcond2}) are solved, is the case with ${\Delta'_h}_\alpha^+ + {\Delta_h}_\alpha^+ = m_\alpha \pi, {\Delta_h}_\alpha^-=\frac{2r_\alpha+1}{2} \pi, {\Delta'_h}_\alpha^-=\frac{2r'_\alpha+1}{2} \pi, P_\alpha S_\alpha=1, r_\alpha+r'_\alpha+m_\alpha+1=S_\alpha$. Nevertheless that this case is completely included in solution (\ref{inverseprescriptions}), it states a specific kind of evolution matrices with ${s_h}_j$ of form:

\begin{eqnarray}
{s_h}_j &=& (-1)^{r_\alpha} e^{i {\Delta_h}_\alpha^+} {\left(
\begin{array}{cc}
-i \beta {j_h}_{-\alpha} & -q i^h {b_h}_{-\alpha}   \\
q {i^*}^h {b_h}_{-\alpha} & i \beta {j_h}_{-\alpha}    
\end{array}
\right) } \\
{{s_h}^{-1}_j} &=& (-1)^{r'_\alpha} e^{i {\Delta'_h}_\alpha^+} {\left(
\begin{array}{cc}
i \beta {j'_h}_{-\alpha} & q i^h {b'_h}_{-\alpha}   \\
-q {i^*}^h {b'_h}_{-\alpha} & -i \beta {j'_h}_{-\alpha}    
\end{array}
\right) } 
\end{eqnarray}

\noindent Note that when $|{j_h}_{-\alpha}|=|{b_h}_{-\alpha}|=\frac{1}{2}$ this sector remembers Hadamard-like gates. 

Nevertheless that prescriptions to get inverses are well defined, they involves sometimes conditions on ${J_h}_\alpha, {J'_h}_\alpha$
which are not possible fulfill in specific experimental designs in terms to find $t', {B'_h}_{-\alpha}$ in terms of $t, {B_h}_{-\alpha}$. Instead, sometimes some of ${J_h},{J'_h},{J_{\{h\}}}_{\pm},{J_{\{h\}}}_{\pm}$ should fulfill strong restrictions (precisely those equivalent to Evolution Loops prescriptions stated before). Last means that inverse evolution for one pulse is not always achievable under general restrictions with other one pulse evolution, despite of inverse is well defined as part of matrices (\ref{mathamiltonian}). 

\subsection{Product closure}

To probe the product closure and to obtain the prescriptions on physical parameters, we can restrict our discussion to sectors again, caring the global matching between exponential factors in (\ref{sector}). Thus, it is necessary probe the matching between ${s'_h}_{j} {s_h}_{j}$ in:

\begin{widetext}
\begin{eqnarray} \label{twopulses}
{s'_h}_{j} {s_h}_{j}={e^{i ({\Delta'_h}_\alpha^+ + {\Delta_h}_\alpha^+)}} \times  
 \left(
\begin{array}{cc}
{{e'_h}_\alpha^\beta}^*{{e_h}_\alpha^\beta}^* - {d'_h}_\alpha {d_h}_\alpha & 
-q i^h ({{e'_h}_\alpha^\beta}^*{d_h}_\alpha + {{e_h}_\alpha^\beta}{d'_h}_\alpha )   \\
q {i^*}^h ({{e'_h}_\alpha^\beta}{d_h}_\alpha + {{e_h}_\alpha^\beta}^*{d'_h}_\alpha ) & 
{{e'_h}_\alpha^\beta}{{e_h}_\alpha^\beta} - {d'_h}_\alpha {d_h}_\alpha    
\end{array}
\right)
\end{eqnarray}
\end{widetext}

\noindent with ${s''_h}_{j}$ with the form in (\ref{sector}). Demonstration is straight but it requires some transformations which we outline briefly. Again, comparing sectors we note that respective equations are compatible only if:

\begin{eqnarray}\label{closcond1}
{{\Delta''_h}_\alpha^+}={{\Delta'_h}_\alpha^+}+{{\Delta_h}_\alpha^+}+2 r_\alpha \pi
\end{eqnarray}

\noindent with $n_\alpha \in {\mathbb Z}$. If this equation fulfills, then only two equations remain, for entries:

\begin{eqnarray}\label{closeeq1}
2,1: && {{e_h}_\alpha^{-\beta}} {{d'_h}_\alpha}  + {{e'_h}_\alpha^{-\beta}} {{d_h}_\alpha} = {{d''_h}_\alpha} \label{closeeq1a}  \\
2,2: && {{e'_h}_\alpha^\beta} {{e_h}_\alpha^\beta} - {{d'_h}_\alpha} {{d_h}_\alpha} = {{e''_h}_\alpha^\beta} \label{closeeq1b} 
\end{eqnarray}

Equation (\ref{closeeq1b}) traduces in two equations for real and imaginary parts respectively:

\begin{eqnarray}
\cos {{\Delta''_h}_\alpha^-} &=& \cos {{\Delta'_h}_\alpha^-} \cos {{\Delta_h}_\alpha^-} - \nonumber \\
&& \quad ({j'_h}_{-\alpha} {j_h}_{-\alpha} + {b'_h}_{-\alpha} {b_h}_{-\alpha}) \cdot \nonumber \\
&& \quad \sin {{\Delta'_h}_\alpha^-} \sin {{\Delta_h}_\alpha^-}   \label{closeeq1bR} \\ \nonumber \\
{j''_h}_{-\alpha} \sin {{\Delta''_h}_\alpha^-} &=& {j'_h}_{-\alpha} \sin {{\Delta'_h}_\alpha^-} \cos {{\Delta_h}_\alpha^-} + \nonumber \\
&& \quad {j_h}_{-\alpha} \cos {{\Delta'_h}_\alpha^-} \sin {{\Delta_h}_\alpha^-} \label{closeeq1bI}
\end{eqnarray}

\noindent then, applying the following transformations $\delta_\pm = {{\Delta'_h}_\alpha^-} \pm {{\Delta_h}_\alpha^-}, \delta j_\pm = \frac{1}{2} ({j_h}_{-\alpha} \pm {j'_h}_{-\alpha}), \delta b_\pm = \frac{1}{2}({b_h}_{-\alpha} \pm {b'_h}_{-\alpha})$, last equations become:

\begin{eqnarray}
\cos {{\Delta''_h}_\alpha^-} &=& (\delta j_+ ^2 + \delta b_+ ^2) \cos \delta_+ + \nonumber \\
&& \quad (\delta j_- ^2 + \delta b_- ^2) \cos \delta_-   \label{closeeq2bR}  \\ \nonumber \\
{j''_h}_{-\alpha} \sin {{\Delta''_h}_\alpha^-} &=&  \delta j_+ \sin \delta_+ + \delta j_- \sin \delta_-  \label{closeeq2bI}
\end{eqnarray}

Following the same process for equation (\ref{closeeq1a}), we obtain that equations for real and imaginary parts become:

\begin{eqnarray}
{b''_h}_{-\alpha} \sin {{\Delta''_h}_\alpha^-} &=&  \delta b_+ \sin \delta_+ + \delta b_- \sin \delta_- \label{closeeq1aR}  \\ \nonumber \\
0 &=& (\delta j_+ \delta b_- - \delta j_- \delta b_+)(\cos \delta_+ - \cos \delta_-)  \nonumber \\ 
&=& ({b_h}_{-\alpha} {j'_h}_{-\alpha} - {b'_h}_{-\alpha} {j_h}_{-\alpha}) \cdot \nonumber \\
&& (\cos \delta_+ - \cos \delta_-) 
\label{closeeq1aI}
\end{eqnarray}

Still, we require that right side in each equation (\ref{closeeq2bR}-\ref{closeeq1aI}) really represent the expression on their left side. More precisely, we need that the following expression becomes one:

\begin{eqnarray}
&& \cos^2 {\Delta''_h}_\alpha^- + ({j''_h}_\alpha^2 + {b''_h}_\alpha^2) \sin^2 {\Delta''_h}_\alpha^- = \quad \quad \nonumber \\ \nonumber \\
&&  \quad = 1-(\delta j_+^2 + \delta b_+^2)(\delta j_-^2 + \delta b_-^2) (\cos \delta_+ - \cos \delta_-)^2 \nonumber \\
&&  \quad = 1-\frac{1-({j_h}_{-\alpha} {j'_h}_{-\alpha} + {b_h}_{-\alpha} {b'_h}_{-\alpha})^2}{4} \cdot \nonumber \\&& \quad \quad \quad \quad (\cos \delta_+ - \cos \delta_-)^2  
\end{eqnarray}

\noindent which, together with (\ref{closeeq1aI}) should be fulfilled. There are two possible solutions: a) $\delta_+ = \pm \delta_- + 2 n \pi, n \in {\mathbb Z}$, which is equivalent to ${\Delta_h}_\alpha^- = n \pi$ or ${\Delta'_h}_\alpha^- = n \pi$, leaving one of them free. This solution is trivial because it implies that one of matrices in the product is proportional to the identity; b) ${j'_h}_{-\alpha} = S_\alpha {j_h}_{-\alpha}, {b'_h}_{-\alpha} = S_\alpha {b_h}_{-\alpha}$, which automatically implies $|S_\alpha|=1$. This situation is trivial when $S_\alpha=1$ and ${J'_{\{h\}}}_{\alpha}, {J_{\{h\}}}_{\alpha}$ do not change because it means constant fields. But if interaction strengths could be manipulated, then it means (for $S_\alpha=\pm 1$):

\begin{eqnarray}\label{condition}
\frac{{J'_{\{h\}}}_\alpha}{{J_{\{h\}}}_\alpha} = \frac{{B'_h}_{-\alpha}}{{B_h}_{-\alpha}}
\end{eqnarray} 

There is one relevant aspect which could be noticed here departing from expressions (\ref{closeeq2bR}-\ref{closeeq1aI}). By considering that all parameters ${j_h}_{-\alpha},{j'_h}_{-\alpha},{b_h}_{-\alpha},{b'_h}_{-\alpha},{{\Delta_h}_\alpha^-},{{\Delta'_h}_\alpha^-}$ are free (until their internal and relative natural restrictions), these right side expressions have a variation between $[-1,1]$ in such way that magnitude of each entry can reach the unit value as before. Nevertheless, by comparison with one pulse case where only multiple integer or multiple semi-integer values are possible, phases in antidiagonal entries can, independently from diagonal ones, reach values in a continuous range inside of $[0,2 \pi)$ but without cover last complete range. This behavior suggest that multiple pulses could extend ${s_h}_j$ coverage into $U(2)$ (or easier, ${s_h}_{j,0}$ coverage into $SU(2)$). 

Summarizing, the prescriptions for product closure in the last terms are:

\begin{eqnarray}
{j''_h}_{-\alpha} &=&  {j_h}_{-\alpha} = S_\alpha {j'_h}_{-\alpha} \nonumber \\
{b''_h}_{-\alpha} &=&  {b_h}_{-\alpha} = S_\alpha {b'_h}_{-\alpha} \nonumber \\
{{\Delta''_h}_\alpha^+} &=& {{\Delta'_h}_\alpha^+}+{{\Delta_h}_\alpha^+}+2 r_\alpha \pi \nonumber \\
{\Delta''_h}_\alpha^- &=&  {{\Delta'_h}_\alpha^-} + S_\alpha {{\Delta_h}_\alpha^-} + 2 r'_\alpha \pi \label{productprescriptions} \\ \nonumber \\
{\rm with:} && S_\alpha  = \pm 1, r_\alpha, r'_\alpha \in {\mathbb Z} \nonumber 
\end{eqnarray}

\noindent showing that each family of evolution matrices (\ref{mathamiltonian}) do not form a group at least that ${j'_h}_{-\alpha} = S_\alpha {j_h}_{-\alpha}, {b'_h}_{-\alpha} = S_\alpha {b_h}_{-\alpha}$ (which automatically implies $|S_\alpha|=1$. This aspect is important because it implies that effect of two o more pulses of magnetic field can not be always replaced with effects achievable by one pulse. These results are consistent with Baker-Campbell-Hausdorff formula for $SU(2)$ reported in \cite{weigert1}:

\begin{eqnarray}\label{expproduct}
{s_h}_{j} {s_h}'_{j}&=& {e^{i ({\Delta_h}_\alpha^+ + {\Delta'_h}_\alpha^+)}} \left( ( 
\cos {{\Delta_h}_\alpha^-} \cos {{\Delta'_h}_\alpha^-}   \right. \nonumber \\
&& \quad \quad - \sin {{\Delta_h}_\alpha^-} \sin {{\Delta'_h}_\alpha^-} {\bf n} \cdot {\bf n}' ) {\mathbb I}_2 \nonumber \\
&& - i (\sin {{\Delta_h}_\alpha^-} \cos {{\Delta'_h}_\alpha^-} {\bf n} + \cos {{\Delta_h}_\alpha^-} \sin {{\Delta'_h}_\alpha^-} {\bf n}' \nonumber \\ 
&& \left. \quad \quad +\sin {{\Delta_h}_\alpha^-} \sin {{\Delta'_h}_\alpha^-} {\bf n} \times {\bf n}') \cdot \boldsymbol{\sigma} \right) \nonumber \\ 
&\equiv& {e^{i {\Delta''_h}_\alpha^+}} \left( \cos {{\Delta''_h}_\alpha^-} {\mathbb I}_2 - i \sin {{\Delta''_h}_\alpha^-} {\bf n''} \cdot {\boldsymbol \sigma} \right) \nonumber \\ \\ \nonumber
{\rm with:} &{\bf n}& = (q {{b_h}_{-\alpha}} \sin \frac{h \pi}{2}, q {{b_h}_{-\alpha}} \cos \frac{h \pi}{2}, \beta {{j_h}_{-\alpha}} ) \nonumber \\
&{\bf n}'& = (q {{b'_h}_{-\alpha}} \sin \frac{h \pi}{2}, q {{b'_h}_{-\alpha}} \cos \frac{h \pi}{2}, \beta {{j'_h}_{-\alpha}} ) \nonumber
\end{eqnarray}

\noindent where $\times$ denotes the vector product. Then, clearly if ${\bf n}'=S_\alpha {\bf n}, |S_\alpha|=1$, both vectors become parallel and prescriptions (\ref{productprescriptions}) are recovered by comparing with (\ref{shjexp}). Note that $U(1)$ factor is abelian and only the part ${{{s''_h}_{j}},0} \in SU(2)$ exhibit a more complex form. Particularly a non vanishing term $\sin {{\Delta_h}_\alpha^-} \sin {{\Delta'_h}_\alpha^-} {\bf n} \times {\bf n}'$ makes non commutative both sectors. 

Coverage of two o more pulses ${s_h}_{j,0}$ into $SU(2)$ could be understood from expression (\ref{expproduct}). Note that
$\cos {{\Delta''_h}_\alpha^-}$ ranges in $[-1,1]$ independently of ${\bf n}'\cdot {\bf n}= ({{b_h}_{-\alpha}} {{b'_h}_{-\alpha}} + {{j_h}_{-\alpha}}{{j'_h}_{-\alpha}})$, when ${{\Delta_h}_\alpha^-}=n \pi$ or ${{\Delta'_h}_\alpha^-}=n \pi$ with $n \in {\mathbb Z}$, leaving other parameter free.  Last is required to ${{s_h}_{j}}_{0}$ have a complete coverage into $SU(2)$. Unfortunately while this variation is nearer from $\pm 1$ ($|{\bf n} \cdot {\bf n}'|=1$, ${{\Delta_h}_\alpha^-}=n \pi$ or ${{\Delta'_h}_\alpha^-}=n \pi$), ${\bf n}''$ have a more limited coverage on unitary sphere as is required to reproduce $SU(2)$.

\subsection{Group structure}

In spite of formulas (\ref{inverseprescriptions}) and (\ref{productprescriptions}), it is clear that $S_\alpha$ can be absorbed as a coefficient of ${\Delta_h}_\alpha^-$. Otherwise, last is equivalent to fix ${j_h}_{-\alpha}$ positive in each group element together with its sign of ${b_h}_{-\alpha}$. Because (\ref{shjexp}), last is equivalent to group those elements by a common unitary vector ${\bf n} = (q {{b_h}_{-\alpha}} \sin \frac{h \pi}{2}, q {{b_h}_{-\alpha}} \cos \frac{h \pi}{2}, \beta {{j_h}_{-\alpha}} )$. In ${{\mathbb S}_h} \subset SU(4)$, each group is characterized by a set of two fixed values $|{{j_h}_{\mp \alpha}}|$ and two signs $s_{{b_h}_{\mp \alpha}}$ correspondingly with signs of ${{b_h}_{\mp \alpha}}$. Thus, each group can be labeled by $\{ |{{j_h}_{\mp \alpha}}| \} , \{ s_{{b_h}_{\mp \alpha}} \}$ and we will denote it by ${{\mathbb S}^*_h}\substack { \{ |{{j_h}_{\mp \alpha}}| \} \\ \{ s_{{b_h}_{\mp \alpha}} \} }$ (being $\pm \alpha$, the first and second sectors in $U_h(t)$ respectively, or $j=1,2$). As extension, we define:

\begin{eqnarray}
{\mathcal C}_h \equiv \bigcup_{\substack {{|{{j_h}_{\pm \alpha}}|} \in [0,1]  \\ s_{{b_h}_{\mp \alpha}} = \pm}} {{\mathbb S}^*_h}\substack { \{ |{{j_h}_{\mp \alpha}}| \} \\ \{ s_{{b_h}_{\mp \alpha}} \} } \subset {{\mathbb S}_h} \subset SU(4)
\end{eqnarray}

Clearly ${\mathcal C}_h=\{U_h(t)\}$ contains all evolution matrices generated by each $U_h(t)$ in (\ref{mathamiltonian}) just grouped by $|{{j_h}_{\mp \alpha}}|$ and $s_{{b_h}_{\mp \alpha}}$ for each sector, which precisely form subgroups in ${{\mathbb S}_h}$. Note that parameters ${\Delta_h}_{\pm \alpha}^\pm$ generate multifolded group elements because periodicity of expressions, it means, several of them generate same group elements in each ${{\mathbb S}_h}$ but they correspond to different evolution dynamics. Same is true for terms $2 r_\alpha \pi, 2 r'_\alpha \pi$ in (\ref{productprescriptions}), which state several alternatives to reproduce an equivalent evolution matrix for two consecutive pulses, more than $r'_\alpha, r_\alpha=0$. Unfortunately for control purposes, this periodicity is not always in the time variable. Thus, in terms of last subsection, each subgroup ${{\mathbb S}^*_h}\substack { \{ |{{j_h}_{\mp \alpha}}| \} \\ \{ s_{{b_h}_{\mp \alpha}} \} }$ (for each $h=1,2,3$) becomes an abelian group. In addition, because that ${{\Delta_h}_\alpha^+} = - {{\Delta_h}_{-\alpha}^+}$ is not a global phase in $U_h(t)$ at least that it reduces to $\pm 1$. Then these relative phase works as an interference frequency in superposition states containing terms of Bell states related with both sectors (separable or partially entangled).

Because rotations explained before, $R_{1,2}(\alpha, \beta, \gamma) \in SU(4)$, transforms elements of each ${\mathbb S}_h$ into other ${\mathbb S}_{h'}$, then subgroups ${{\mathbb S}^*_1}\substack { \{ |{{j_1}_{\pm}}| \} \\ \{ s_{{b_1}_{\pm}} \} }, {{\mathbb S}^*_2}\substack { \{ |{{j_2}_{\mp}}| \} \\ \{ s_{{b_2}_{\mp}} \} }, {{\mathbb S}^*_3}\substack { \{ |{{j_3}_{\pm}}| \} \\ \{ s_{{b_3}_{\pm}} \} }$ are clearly isomorphic \cite{cornwell1} between them under these rotations. Moreover, if ${\mathcal T}$ is a transformation with properties:

\begin{eqnarray}\label{transformation}
{\mathcal T}:{s_h}_j \in {{\mathbb S}^*_h}\substack { \{ |{{j_h}_{\mp \alpha}}| \} \\ \{ s_{{b_h}_{\mp \alpha}} \} } &\rightarrow& {{s}^t_h}_j \in {{\mathbb S}^*_h}\substack { \{ |{{j'_h}_{\mp \alpha}}| \} \\ \{ s_{{b'_h}_{\mp \alpha}} \} } \nonumber \\
{\mathcal T}({s'_h}_j {s_h}_j)&=&{\mathcal T}({s'_h}_j) {\mathcal T}({s_h}_j)
\end{eqnarray}

\noindent then, in spite of (\ref{productprescriptions}), ${{\mathbb S}^*_h}\substack { \{ |{{j_h}_{\mp \alpha}}| \} \\ \{ s_{{b_h}_{\mp \alpha}} \} }$ and $ {{\mathbb S}^*_h}\substack { \{ |{{j'_h}_{\mp \alpha}}| \} \\ \{ s_{{b'_h}_{\mp \alpha}} \} }$ become isomorphic because product structure is preserved:

\begin{eqnarray}
{{s''}^t_h}_j={\mathcal T}({s''_h}_j)={\mathcal T}({s'_h}_j {s_h}_j)={\mathcal T}({s'_h}_j) {\mathcal T}({s_h}_j)={{s'}^t_h}_j {{s}^t_h}_j \nonumber \\
\end{eqnarray}

Note that in spite (\ref{closeeq1aI}), if prescription (\ref{condition}) is fulfilled, then ${s'_h}_{j},{s_h}_{j}$ will commute at sector level (yet, prescription should be fulfilled for $\pm \alpha$ in order that ${U'_h}(t),{U_h}(t)$ commute). Nevertheless, is easy note that (\ref{condition}) automatically implies ${j'_h}_{-\alpha}=S_\alpha {j_h}_{-\alpha}, {b'_h}_{-\alpha}=S_\alpha {b_h}_{-\alpha}, S_\alpha=\pm 1$. Thus, ${\mathcal C}_h$ contains all sets of single commutative (not between sets) evolution matrices  of form ${U_h}(t)$ in (\ref{mathamiltonian}).  

\begin{figure}[pb] 
\centerline{\psfig{width=3in,file=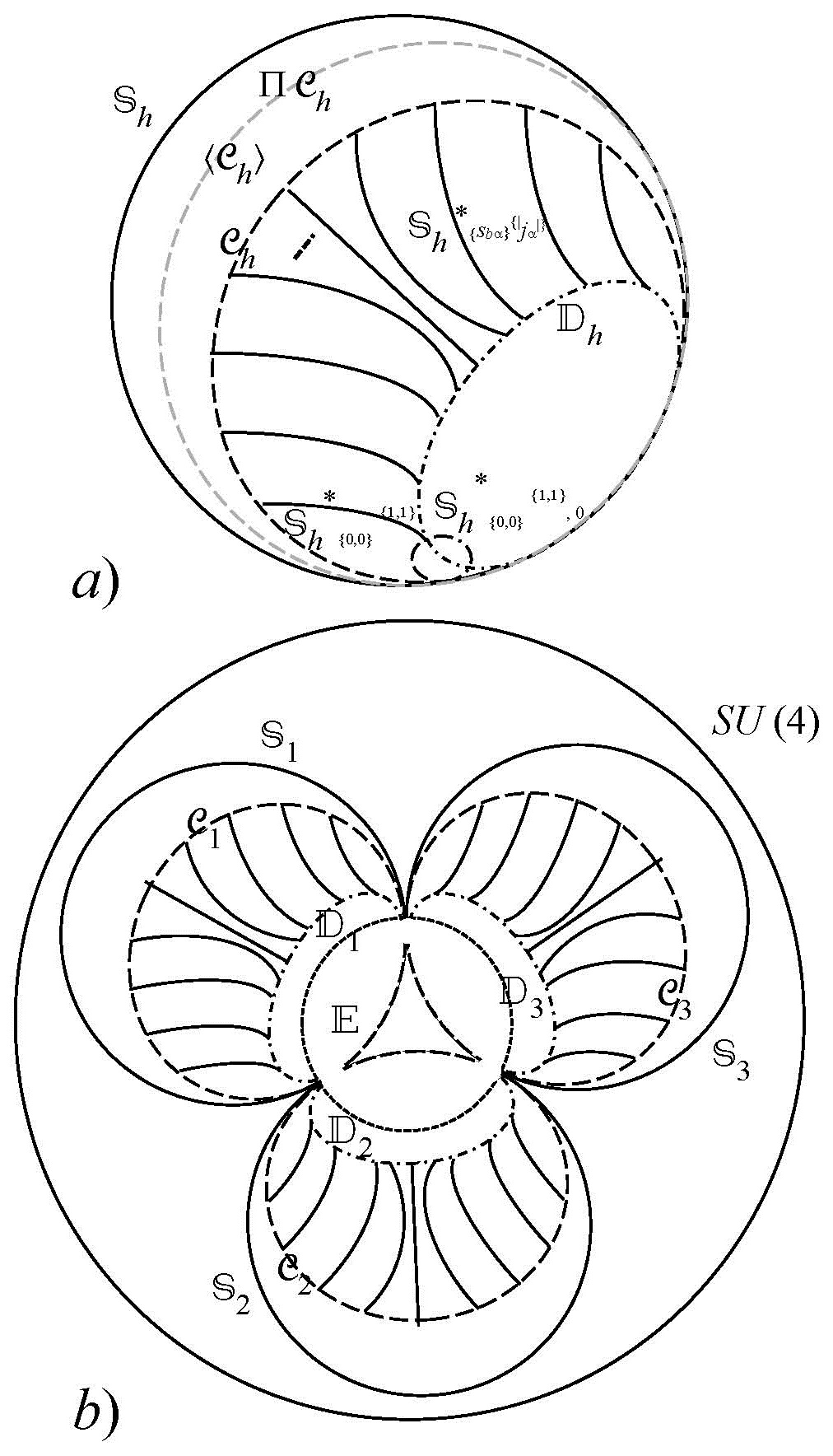}}
\vspace*{8pt}
\caption{Schematic representation of each ${{\mathbb S}_h}$ for $h=1,2,3$ in $SU(4)$. a) Structure of ${{\mathbb S}_h}$ with an infinite number of sections, each one containing a different subgroup ${{{\mathbb S}^*_h}}^{\{ |{{j_h}_{\mp \alpha}}| \}}_{\{ s_{{b_h}_{\mp \alpha}} \}}$ parametrized in $[0,1]^{\times 2} \times \{\pm\}^{\times 2}$ and covering ${\mathcal C}_h$ (dotted oval region). Sections extends on oval white region containing ${\mathbb D}_h$ (indicated with the inner dotted boundary) in which all subgroups ${{{\mathbb S}^*_h}}^{\{ |{{j_h}_{\mp \alpha}}| \}}_{\{ s_{{b_h}_{\mp \alpha}} \}}$ share elements. Note specially the subgroup ${{{\mathbb S}^*_h}}^{\{ 1,1 \}}_{\{ 0,0 \}}$ and their subgroup ${{{\mathbb S}^*_h}}^{\{ 1,1 \}}_{\{ 0,0 \},0}$. ${\mathcal C}_h$ does not cover ${{\mathbb S}_h}$. Uncovered region (semi lunar region in white) could be divided in two regions by the gray circular dotted line: inner region is the group $\left< {\mathcal C}_h \right>$, which contains all finite products of elements in ${\mathcal C}_h$; and outer one contains the elements in ${\mathbb S}_h$ not achievable by any finite product of elements in ${\mathcal C}_h$. b) Combination of three subgroups ${{\mathbb S}_h}$ being contained in $SU(4)$, which share elements in the central circular region ${\mathbb E}$.} 
\label{fig6}
\end{figure}

In the further discussion, the following additional sets and subgroups are notable in terms of group theory, which will be shown in Figure \ref{fig6}. The first one are the subgroups in ${{\mathbb S}^*_h}\substack { \{ |{{j_h}_{\mp \alpha}}| \} \\ \{ s_{{b_h}_{\mp \alpha}} \} }$:

\begin{eqnarray}
{{{\mathbb S}^*_h}\substack { \{ |{{j_h}_{\mp \alpha}}| \} \\ \{ s_{{b_h}_{\mp \alpha}} \} }}_{,0} \equiv \{ U_h (t) \in {{\mathbb S}^*_h}\substack { \{ |{{j_h}_{\mp \alpha}}| \} \\ \{ s_{{b_h}_{\mp \alpha}} \} } | {{\Delta_h}_\alpha^+} = 2 n \pi, n \in {\mathbb Z} \} \nonumber \\
\end{eqnarray}

\noindent which clearly correspond to those where sectors phases $e^{\pm i {{\Delta_h}_\alpha^+}}$ have been removed. This subgroup contains the ${{s_h}_{j}}_{,0}$ matrices defined in (\ref{shjexp}). In particular, ${{{\mathbb S}^*_h}\substack { \{ 1,1 \} \\ \{ 0,0 \} }}_{,0}$ is the diagonal matrix whose sectors have only opposite phases in each entry (taking $s_{{b_h}_{\mp \alpha}}=0$ as exception to our notation in (\ref{graldiag2}) because $|{j_h}_{\pm \alpha}|=1 \rightarrow |{b_h}_{\pm \alpha}|=0$). Matrices in this subgroup are responsible to introduce relative phases between Bell states associated with the same sector. Other relevant subgroup is:

\begin{eqnarray}
{{\mathbb D}_h} \equiv \{ U_h (t) \in {{\mathbb S}^*_h}\substack { \{ |{{j_h}_{\mp \alpha}}| \} \\ \{ s_{{b_h}_{\mp \alpha}} \} } | {{\Delta_h}_{\pm \alpha}^-} = n_{\pm \alpha} \pi, n_{\pm \alpha} \in {\mathbb Z} \} \nonumber \\
\end{eqnarray} 

\noindent they correspond precisely to matrices ${{\mathcal D}_h^\phi}$ presented before, responsible to introduce relative phases between Bell states associated with different sectors. These matrices are contained in any ${{{\mathbb S}^*_h}\substack { \{ |{{j_h}_{\mp \alpha}}| \} \\ \{ s_{{b_h}_{\mp \alpha}} \} }}$ so they commute with any matrix in ${\mathcal C}_h$ and inclusively in ${{\mathbb S}_h}$. Thus, in terms of group theory, ${{\mathbb D}_h}$ is a normal group \cite{cornwell1} of ${{\mathbb S}_h}$: ${{\mathbb D}_h} \vartriangleleft {{\mathbb S}_h}$. In a related sense, given a subgroup $A$ in $G$, the left cosets (similarly for right cosets) are defined as sets:

\begin{eqnarray}
g A \equiv \{g a | g \in G, a \in A \}
\end{eqnarray} 

\noindent with this, we can set a result of group theory. Because ${{\mathbb D}_h}$ is a normal group, then the set denoted by $G / {{\mathbb D}_h}$ of left (or right) cosets on bigger group $G$ (possible groups represented by $G$ are ${{{\mathbb S}^*_h}\substack { \{ 1,1 \} \\ \{ 0,0 \} }}_{,0}, {{{\mathbb S}^*_h}\substack { \{ |{{j_h}_{\mp \alpha}}| \} \\ \{ s_{{b_h}_{\mp \alpha}} \} }}, {{\mathbb S}_h}$ or inclusively $SU(4)$) which contains ${{\mathbb D}_h}$, is the quotient group \cite{cornwell1} of $G$ by ${{\mathbb D}_h}$. It can be understood as a homomorphism on $G$ and it is easy note that ${{\mathbb D}_h}$ is the kernel of this homomorphism (the set of elements in $G$ which maps on ${{\mathbb D}_h}$). 

If $A, B$ are subgroups in $G$, then we define their product \cite{cornwell1} as:

\begin{eqnarray}
AB = \{ ab | a \in A, b \in B \}
\end{eqnarray}

\noindent which is not necessarily a group at least one of them be a normal group which is a result well known in group theory \cite{cornwell1}. Then should be clear that:

\begin{eqnarray}
 {{{\mathbb S}^*_h}\substack { \{ |{{j_h}_{\mp \alpha}}| \} \\ \{ s_{{b_h}_{\mp \alpha}} \} }} = {{\mathbb D}_h} {{{\mathbb S}^*_h}\substack { \{ |{{j_h}_{\mp \alpha}}| \} \\ \{ s_{{b_h}_{\mp \alpha}} \} }}_{,0}={{{\mathbb S}^*_h}\substack { \{ |{{j_h}_{\mp \alpha}}| \} \\ \{ s_{{b_h}_{\mp \alpha}} \} }}_{,0} {{\mathbb D}_h} 
\end{eqnarray}

Under this circumstances, this set represents how ${{\mathbb D}_h}$ can be understood as a merely factor on the whole structure of group $G$. Summarizing, in group theory language, ${{{\mathbb S}^*_h}\substack { \{ |{{j_h}_{\mp \alpha}}| \} \\ \{ s_{{b_h}_{\mp \alpha}} \} }} = {{\mathbb D}_h} \times {{{\mathbb S}^*_h}\substack { \{ |{{j_h}_{\mp \alpha}}| \} \\ \{ s_{{b_h}_{\mp \alpha}} \} }}_{,0}$ is a direct product of these subgroups \cite{cornwell1}, which physically means that evolution operators obtained with one magnetic pulse are equivalent to two pulses, one to provide sector phases and other to reproduce the remaining dynamics free of sector phases.

Finally, the subgroup:

\begin{eqnarray}
{{\mathbb E}} \equiv \{e^{i \phi} {\mathbb I}_4 | \phi \in [0,2\pi) \} \subset \bigcap_{\substack {h=1,2,3}} {{\mathbb S}_h}
\end{eqnarray}

\noindent can be understood as responsible to introduce global phases in quantum states under interaction (\ref{hamiltonian}) as difference with those phases introduced by ${{{\mathbb S}^*_h}\substack { \{ 1,1 \} \\ \{0,0\} }}_{,0}$ and ${\mathcal D}_h$. Clearly, as ${{{\mathbb S}^*_h}\substack { \{ |{{j_h}_{\mp \alpha}}| \} \\ \{ s_{{b_h}_{\mp \alpha}} \} }}_{,0}$ and ${\mathbb D}_h$, ${{\mathbb E}}$ is isomorphic to $U(1)$. Thus, similarly with ${{\mathbb D}_h}$, $G / {{\mathbb E}}$ is a quotient group, representing how a global phase is a factor group $U(1)$ in the structure of each possible $G$ in this analysis (${{{\mathbb S}^*_h}\substack { \{ 1,1 \} \\ \{ 0,0 \} }}_{,0}, {{{\mathbb S}^*_h}\substack { \{ |{{j_h}_{\mp \alpha}}| \} \\ \{ s_{{b_h}_{\mp \alpha}} \} }}, {{\mathbb S}_h}$ or $SU(4)$).

Figure \ref{fig6} depicts schematically the relations between before sets and subgroups in $SU(4)$. Despite their representations (\ref{mathamiltonian}), each ${{\mathbb S}_h}$ (Figure \ref{fig6}a) has same group structure as a result of their isomorphic relations via rotations $R_{1,2}(\alpha, \beta, \gamma)$. Inside, there are an infinite number of subgroups ${{{\mathbb S}^*_h}\substack { \{ |{{j_h}_{\mp \alpha}}| \} \\ \{ s_{{b_h}_{\mp \alpha}} \} }}$ parametrized on $[0,1]^{\times 2} \times \{\pm\}^{\times 2}$, which cover ${\mathcal C}_h$. Subgroup ${\mathbb D}_h$ contains the common elements of them. Subgroups ${{{\mathbb S}^*_h}\substack { \{ |{{j_h}_{\mp \alpha}}| \} \\ \{ s_{{b_h}_{\mp \alpha}} \} }}_{,0}$ (not indicated in figure) are each one in their corresponding ${{{\mathbb S}^*_h}\substack { \{ |{{j_h}_{\mp \alpha}}| \} \\ \{ s_{{b_h}_{\mp \alpha}} \} }}$. When three groups ${{\mathbb S}_h} \subset SU(4)$ are combined (Figure \ref{fig6}b), subgroup ${\mathbb E}$ contains their only common elements. Thus, oval regions with dotted boundary, ${\mathcal C}_h$, in each ${{\mathbb S}_h}$ represent the evolutions achievable with only one pulse of magnetic field. Instead, white semi lunar regions $\Pi {\mathcal C}_h ={{\mathbb S}_h} \diagdown {\mathcal C}_h$ are complementary evolutions which acts 'locally' and unitarily in each matrix sector but not able to be generated in just one pulse. 

Other important result in group theory is that given a set of groups, by example ${\mathcal C}_h \in {{\mathbb S}_h}$, then their generated subgroup is defined as all finite products of elements and/or their inverses:

\begin{eqnarray}
\left< {\mathcal C}_h \right> \equiv \{ u=a_1 a_2 \hdots a_n | n \in {\mathbb Z^+}, a_i \in {\mathcal C}_h \cup {\mathcal C}^{-1}_h \}
\end{eqnarray}

\noindent where ${\mathcal C}^{-1}_h \equiv \{ a_i^{-1} | a_i \in {\mathcal C}_h \}$ (in this case ${\mathcal C}_h = {\mathcal C}^{-1}_h$). Then, $\left< {\mathcal C}_h \right>$ have the property of being a group, which is easily demonstrable departing from group conditions. It means that finite products of elements in ${\mathcal C}_h$ extends their coverage into ${{\mathbb S}_h}$ forming a group \cite{cornwell1}, but it is not necessarily whole ${{\mathbb S}_h}$ (this group is depicted in Figure \ref{fig6} with a gray dotted circular line): ${\mathcal C}_h \subset \left< {\mathcal C}_h \right> \subset {\mathbb S}_h$. Note that $\left< {\mathcal C}_h \right>$ is generated by the minimum set:

\begin{eqnarray}
{\mathcal F}_h \equiv {{\mathbb D}_h} \cup ( \bigcup_{\substack {{|{{j_h}_{\pm \alpha}}|} \in [0,1]  \\ s_{{b_h}_{\mp \alpha}} = \pm}} {{\mathbb S}^*_h}\substack { \{ |{{j_h}_{\mp \alpha}}| \} \\ \{ s_{{b_h}_{\mp \alpha}} \} }_{,0} )
\end{eqnarray}

\noindent thus, in terms of group theory, $\left< {\mathcal C}_h \right>$ is a free group on ${\mathcal F}_h$. In these terms, $\Pi {\mathcal C}_h \diagdown \left< {\mathcal C}_h \right>$ contains all possible elements in ${{\mathbb S}_h}$ which are not achievable by any finite combination of interaction pulses (\ref{hamiltonian}) for a fixed $h$. Elements in this last set have the following property: $p_i p_j = s_k, \forall p_i,p_j \in \Pi {\mathcal C}_h \diagdown \left< {\mathcal C}_h \right>$ with $s_k \in \left< {\mathcal C}_h \right>$. It follows that a) $\left< {\mathcal C}_h \right>$ self-contains their inverses; b) for any cosets generated by $p \in \Pi {\mathcal C}_h \diagdown \left< {\mathcal C}_h \right>$, then $p \in p \left< {\mathcal C}_h \right>$; and c) by hypothesis, no one element of $\Pi {\mathcal C}_h \diagdown \left< {\mathcal C}_h \right>$, can be obtained as a finite product of elements of $\left< {\mathcal C}_h \right>$. As a corollary, the product of an even number of elements $p_i$ of  $\Pi {\mathcal C}_h \diagdown \left< {\mathcal C}_h \right>$ becomes element of $\left< {\mathcal C}_h \right>$, while the product of an odd number remains in this set.

Note that formula (\ref{expproduct}) is useful and easy to determine how two pulses ${s'_h}_{j}, {s_h}_{j}$ with ${\bf n'},{\bf n}$ not parallel generate elements outside of ${\mathcal C}_h$. Because form of ${\bf n}$ in (\ref{expproduct}), it shows too that for ${s_h}_{j}^a, {s_h}_{j}^b$ with given ${j_h}^a_{\mp \alpha}, {j_h}^b_{\mp \alpha}$ and $s_{{b_h}_{\mp \alpha}}^a,s_{{b_h}_{\mp \alpha}}^b$ respectively, then it is not possible to find correspondingly a ${s_h}_{j}^c$ which fulfill ${s_h}_{j}^a {s_h}_{j}^b = {s_h}_{j}^c$ at least ${\bf n}^a,{\bf n}^b$ will be parallel and then ${\bf n}^b$ (owning to same subgroup). Then, finite products in ${\mathcal C}_h$ are not trivial because they can not easily be rearranged to simplify their structure, except in the case when two pulses belong to the same group ${{\mathbb S}^*_h}\substack { \{ |{{j_h}_{\mp \alpha}}| \} \\ \{ s_{{b_h}_{\mp \alpha}} \}}$. 

\begin{figure}[pb] 
\centerline{\psfig{width=2.8in,file=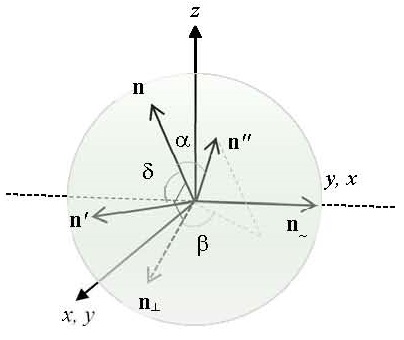}}
\vspace*{8pt}
\caption{Generation of an arbitrary element of $SU(2)$ for sector in elements of $U_h(t)$ with two pulses. Basis ${\bf n}, {\bf n}_\perp, {\bf n}_\backsim$ let express general components of ${\bf n}''$, letting to obtain solutions for it and ${\Delta''_h}_{\pm \alpha}^-$ in terms of ${{\Delta_h}_{\pm \alpha}^-}, {{\Delta'_h}_{\pm \alpha}^-}, \delta$.}
\label{fig8}
\end{figure}

As $SU(2)$, each set of sectors in a given ${{{\mathbb S}^*_h}\substack { \{ |{{j_h}_{\mp \alpha}}| \} \\ \{ s_{{b_h}_{\mp \alpha}} \} }}$ group is a Lie group with ${\Delta_h}_\alpha^-$ as parameter. A result in Lie group theory is that every element of the connected subgroup of any linear Lie group can be expressed as a finite product of exponentials of its real linear Lie algebra \cite{cornwell1,robart1}. So, if sectors of elements in $\left< {\mathcal C}_h \right>$ are connected, it will imply that this last group is $SU(2)$ really. We can analyze connectivity with help of formula (\ref{expproduct}). Clearly elements in ${{{\mathbb S}^*_h}\substack { \{ |{{j_h}_{\mp \alpha}}| \} \\ \{ s_{{b_h}_{\mp \alpha}} \} }}$ have ${\bf n}$ restricted to planes $1-3$ or $2-3$ (or $x-z$,$y-z$ but not associated with physical directions) depending of $h$ parity. Thus, ${\bf n} \times {\bf n}'$ in (\ref{expproduct}) is orthogonal to these vectors. With this, we can define the following orthonormal vector basis (Figure \ref{fig8}):

\begin{eqnarray}
&{\bf n} \nonumber \\
&{\bf n}_\perp = {\bf n}'- \cos \delta \quad {\bf n} \label{nperp}  \label{northo} \\
&{\bf n}_\backsim = \csc \delta \quad {\bf n} \times {\bf n}' \nonumber 
\end{eqnarray}

\noindent where $\cos \delta = {\bf n} \cdot {\bf n}'$. Then, it is possible to express ${\bf n}''$ in (\ref{expproduct}) in a spherical system of coordinates as:

\begin{eqnarray}
{\bf n}'' =  \cos \alpha {\bf n} + \sin \alpha \cos \beta {\bf n}_\perp + \sin \alpha \sin \beta {\bf n}_\backsim
\end{eqnarray}

\noindent becoming when we solve:

\begin{eqnarray}
{\Delta_h}_{\pm \alpha}^- &=& \beta + n \pi , n \in {\mathbb Z} \nonumber \\
\sin {\Delta'_h}_{\pm \alpha}^- &=& \sin {\Delta''_h}_{\pm \alpha}^- \sin \alpha \csc \delta  \\
\cot \delta &=& \frac{\sin {\Delta''_h}_{\pm \alpha}^- \cos \alpha \cos \beta + \cos {\Delta''_h}_{\pm \alpha}^- \sin \beta }{\sin {\Delta''_h}_{\pm \alpha}^- \sin \alpha} \nonumber 
\end{eqnarray}

\noindent which solves the desired output element in terms of input parameters. Right side expression for $\sin {\Delta'_h}_{\pm \alpha}^-$ ranges between $[-1,1]$ as is required (it can be shown by combining two last expressions and then obtaining their extreme values). A brief analysis shows that $\cot \delta$ ranges in ${\mathbb R}$. Then, two pulses in different groups ${{{\mathbb S}^*_h}\substack { \{ |{{j_h}_{\mp \alpha}}| \} \\ \{ s_{{b_h}_{\mp \alpha}} \} }}$ can be adequately selected to reproduce a general element in $SU(2)$ (still, cases when $\theta, \alpha= n \pi, n \in {\mathbb Z}$ can be obtained as limit cases; some of them are cases discussed before for diagonal or antidiagonal forms). Clearly, there are several solutions because they depend only on $\delta$, the angle between original vectors ${\bf n}$ and ${\bf n}'$, more than representation of specific vectors being considered. With this, $\Pi {\mathcal C}_h \diagdown \left< {\mathcal C}_h \right>=\emptyset$. This result is useful because states that elements in $SU(2)$  are products of sectors in $U_h(t)$. With this, each element in ${{\mathbb S}_h}$ can be obtained in a finite number of pulses, each one belonging to ${{{\mathbb S}^*_h}\substack { \{ |{{j_h}_{\mp \alpha}}| \} \\ \{ s_{{b_h}_{\mp \alpha}} \} }}$ groups.

This analysis state algebraic properties for solutions (\ref{mathamiltonian}) of Ising interaction model (\ref{hamiltonian}) being considered. This perspective lets combine them for different purposes in control, gates and states design between others.

\section{Conclusions}

Physical systems as molecules to set databases \cite{john1}, magnetic molecular clusters and dielectric nanometer-size single domain to set quantum information processing \cite{klo1}, spins in quantum dots formed in GaAs heterostructures, nanowire-based quantum dots or self-assembled quantum dots as suitable qubits \cite{loss1,ima1}, are some examples of physical systems on which spin control has been experimented. Part of idea in those systems is to have sufficient ability to have single resources on which set quantum computation and quantum information processing in terms of before quantum computer models \cite{john1} with programmable spin–spin couplings as some of shown here. It is clear through these examples that different physical systems can converge on very similar kind of interactions which require deep analysis of their models to be experimentally exploited. In addition, it is clear that as experimental and technology advance, then more fine control has been applied to control paths, stable equilibrium, confinement and quantum states of course. Then, models with extended parameters of control should be analyzed because them could bring a better performance in the quantum states control arena. These theoretical developments sooner or later meet with experimental developments in order to become useful in quantum engineering.

Nevertheless that models with several couplings including more that two qubits at time, these trends are studied to improve some proposals of superdense coding, multi-entangled processing, quantum walks and other which require it. Despite control is being improved, it has been clear that decoherence in a multiqubits deployments increases easily with their parts number, so manipulate lots of qubits coordinately becomes difficult in a growing array \cite{klo1}. Thus, alternative well controlled developments based on a few quantum qubits at time should be developed to implement quantum algorithms being constructed specifically to these kind of systems while multipartite control is better understood and improved. Circuit-gate model of quantum computers could be based in great extent on bipartite systems when ancilla qubits are used. 

Clearly, extensions to a few more qubits will be necessary still because such quantum stuff requires a system of several qubits to make some task efficiently and the main materials based technology known for that is magnetic. The most of them exploits Ising interactions with different approaches \cite{john1}, together with control on quantum states and in particular with entanglement control, a milestone in all almost these researches. Analysis presented in this work states some properties which generalize some restricted models used in several approaches and experimental setups. In this sense, three dimensional model can reduce to simplified models but brings potentially extensions in those models and technologies.

Circuit-gate model was the first approach to quantum computation, nevertheless, quantum annealing \cite{kado1} or measurement-based quantum computation \cite{briegel1} are alternatives which use magnetic systems approached by Ising interactions to manage a planned and controlled quantum state manipulation. By example, \cite{ying1} has proposed a scheme to simulate the Ising model and preserve the maximum entangled states (Bell states) in cavity quantum electrodynamics (QED) driven by a classical field with large detuning. On them, several applied problems has been exhibited as the goal of (these technologies (pattern matching, folding proteins, an other particular NP-complete problems \cite{john1}).

In these directions, solution for Ising model presented here (\ref{hamiltonian}) can be applied for more controlled situations including more than three freedom degrees. Magneto-optic traps an QED cavities could tentatively manage control of position and contain particles, ions or molecules leaving still three directions for spin-spin coupling. Still, not all of specific effects need use three dimensional freedom degrees but other developments can extent their external and internal dynamics into three dimensions (by example, \cite{mich1}, has been reported three dimensional tracking of quantum dots).

Future work for model and solutions (\ref{mathamiltonian}) presented could be based in different research lines. One proposal is to grown the analysis to control of elementary pieces to set adequate resources in circuit-gate quantum computer in terms of (\ref{evloop}-\ref{exch}). Control procedures as Evolution Loops or Exchange Operations states a basic structure and language of manipulation to maintain or transform qubits selectively. This basic language lets translate the circuit-gate quantum computation algorithms into physical operations based on realistic systems. After of control analysis to set stable and recoverable quantum resources, other possible extensions are based on statement of a computer grammar based on those resources (as states Figure \ref{fig0} and structure depicted in terms of Bloch sphere to pairs of Bell states. An outstanding aspect here was the introduction of non local basis to depicts dynamics, which uncover the regular forms in (\ref{mathamiltonian}) with well understood group structures. It is possible that for models including more qubits, this structure could be maintained in terms of an adequate basis of maximal entangled states as in (\ref{evop2}) for two qubits. In terms of these expressions, our interaction appear as operate almost independently on pairs of maximal entangled states.

Another extension is to exploit possibilities for two pulses expression (\ref{twopulses}) or in general a finite product of pulses, which extend group dominion on ${\mathbb S}_h$. Still, combination of those operators for different values of $h$ should be studied to state its coverage on $SU(4)$.   

In this line of research, the analysis of behavior with finite temperature based on matrix density is compulsory to consider decoherence effects. At same time, error correction analysis is necessary in procedures which emerge of present model, based on error factors (as magnetic field, knowledge and control of interaction strengths, time, etc.). In our approach, of course improvements should be generated through to alternative continuous pulses. Rectangular pulses are easy to manage theoretically but reality is that they are experimentally few practical because their discontinuity and associated resonant effects.

Just as control development advances, more complex models can be experimented. Nuclear magnetic resonance, Quantum dots and Electrons in silicon lattices have been the most successful systems in implementing quantum algorithms based on their coherence and stability. If quantum computation could be based on entangled basis as a grammar, despite the actual complications about their maintenance, it could be a more understandable language to generate a programming basis because they are adapted to physical systems where they are set up. In that terms, proposals as here presented increase potentially their value.


\small  


\begin{thebibliography}{5}

\bibitem{vneumann1}
von Neumann, J., {\it Mathematische Grundlagen der Quantenmechanic}, (Springer, Berlin, 1932).

\bibitem{schrod1} 
E. Schr\"odinger and M. Born, Proc. Cambridge Phil. Soc. {\bf 31} (4), 555 (1935).

\bibitem{schrod2} 
E. Schr\"odinger, Naturwissenschften {\bf 23}, 807 (1935).

\bibitem{einstein1} 
A. Einstein, B. Podolsky, and N. Rosen, Phys. Rev. {\bf 47}, 777 (1935).

\bibitem{schrod3}
E. Schr\"odinger and P. A. M. Dirac, Mathematical Proceedings of the Cambridge Philosophical Society {\bf32} (3), 446 (1936). 

\bibitem{jozsa1}
R. Jozsa and N. Linden, Proc. Royal Soc. A: Mathematical, Physical and Engineering Sciences {\bf 459} (2036), 2011 (2002).

\bibitem{jozsa2}
R. Jozsa, {\it Entanglement and quantum computation}, e-print quant-ph/9707034.

\bibitem{bennet1}
C. H. Benett, D. P. DiVincenzo, J. A. Smolin and W. K. Wooters, Phys. Rev. A {\bf 54}, 3824 (1996).

\bibitem{feyn1}
R. P. Feynman, Int. J. Theor. Phys. {\bf 21}, 467 (1982).

\bibitem{deutsch1}
D. Deutsch, Proc. R. Soc. London, Ser. A {\bf 400}, 97 (1985).

\bibitem{steane1}
A. Steane, Phys. Rev. Lett {\bf 77}, 793 (1996).

\bibitem{bennet2} 
C. H. Bennett and G. Brassard, Proc. IEEE Intl. Conf. on Comp. {\bf 175}, (1984).

\bibitem{ekert1}
Ekert, A., Phys. Rev. Lett {\bf 67}, 661 (1991).

\bibitem{bennet3}
C. H. Bennett and S. J. Wiesner, Phys. Rev. Lett {\bf 69}, 2881 (1992).

\bibitem{bennet4}
C. H. Bennett, G. Brassard, C. Crepeau, R. Jozsa, A. Peres and W. K. Wootters, Phys. Rev. Lett {\bf 70}, 1895 (1993).

\bibitem{ising1}
E. Ising, Z. Phys. {\bf 31}, 253 (1925).

\bibitem{brush1}
S.G. Brush, Rev. Mod. Phys. {\bf 39}, 883 (1967).

\bibitem{baxter1}
R.J. Baxter, {\it Exactly solved models in statistical mechanics}, (Acad. Press, 1982).

\bibitem{nielsen2}
M. A. Nielsen, Ph. D. Thesis, University of New Mexico, 1998; see also LANL e-print: quant-ph/0011036

\bibitem{kiri1}
I. L. Kirilyuk and S. V. Prants, in Proceedings of 2nd International Conference in Control of Oscillations and Chaos (IEEE,
New York, 2000), Vol. 2, p. 369.

\bibitem{meng1}
Q. T. Meng, G. H. Yang, and K. L. Han, Int. J. Quantum Chem.
95, 30 (2003).

\bibitem{terzis1}
A.F. Terzis and E. Paspalakis, Phys. Lett. A {\bf 333}, 438 (2004).

\bibitem{stelma1}
P. Stelmachovic and V. Buzek, Phys. Rev. A {\bf 70}, 032313 (2004).

\bibitem{novo1}
J. Novotny, M. Stefa˜nak, T. Kiss, and I. Jex, J. Phys. A: Math. Gen. {\bf 38}, 9087 (2005).

\bibitem{recher1}
P. Recher and D. Loss, in Proceedings of SPINTRONICS 2001: International Conference on Novel Aspects of Spin-Polarized Transport and Spin Dynamics (Springer, New York, 2002), Vol. 15, p. 49.

\bibitem{saraga1}
D. S. Saraga, B. L. Altshuler, D. Loss, and R. M. Westervelt, Phys. Rev. Lett. {\bf 92}, 246803 (2004).

\bibitem{kopp1}
F. H. L. Koppens, C. Buizert, K.J. Tielrooij, I. T. Vink, K. C. Nowack, T. Meunier, L. P. Kouwenhoven1 and L. M. K. Vandersypen, Nature {\bf 442} (7104), 766 (2006).

\bibitem{vincenzo1}
D. P. DiVincenzo, in Mesoscopic Electron Transport, of NATO Advanced Study Institute, Series E: Applied Sciences (Kluwer, Dordrecht, 1997), Vol. 345, p. 657.

\bibitem{berman1}
G.P. Berman, G. D. Doolen, G.V. L\'opez, and V. I. Tsifrinovich, {\it Generalized Quantum Control-Not Gate in Two-Spin Ising System}, quant-ph/9802013v1.

\bibitem{wang2}
X. Wang, Phys. Lett. A {\bf 281}, 101 (2001).

\bibitem{aless1}
D. D’alessandro, {\it Introduction to Quantum Control and Dynamics} (Chapman Hall Applied Mathematics Nonlinear Science 2007).

\bibitem{meek1}
Meekhof, D. M., C. Monroe, B. E. King, W. M. Itano, and D. J. Wineland, Phys. Rev. Lett {\bf 76}, 1796 (1996).

\bibitem{rai1}
Raimond, J. M., M. Brune, and S. Haroche, Rev. Mod. Phys. {\bf 73}, 565 (2001).

\bibitem{ange1}
D. G. Angelakis, M. F. Santos and S. Bose, Physical Review A {\bf 76} (03), 1805 (2007).

\bibitem{branczyk1}
A. M. Bra\'{n}czyk, P. E. M. F. Mendon\c{c}a, A. Gilchrist, A. C. Doherty and S. D. Bartlett,Phys. Rev. A {\bf 75}, 012329 (2007).

\bibitem{xi1}
Z. Xi and G. Jin, Int. J. Quant. Info. {\bf 5}, 857 (2007).

\bibitem{delgado1}
F. Delgado, (2010), Rev. Mex. Fis. {\bf 56} (1) 30.

\bibitem{delgado2}
F. Delgado, (2010), Phys. Rev. A {\bf 81}, 042317.

\bibitem{kamta1}
G. L. Kamta and A. F. Starace, Phys. Rev. Lett. {\bf 87}, 017901 (2001).

\bibitem{sun1}
Y. Sun, Y. Chen and H. Chen, Phys. Rev. A {\bf 68}, 044301 (2003).

\bibitem{zhou1}
L. Zhou, H. S. Song, Y. Q. Guo and C. Li, Phys. Rev. A {\bf 64}, 042302 (2001).

\bibitem{gunlycke1}
D. Gunlycke, V. M. Kendon, V. Vedral and S. Bose, Phys. Rev. A {\bf 64}, 042302 (2001).

\bibitem{wott1}
W. K. Wotters, Quant. Inf. Comp. {\bf 1}, 27 (2001).

\bibitem{mielnik1}
B. Mielnik, J. Math. Phys. {\bf 27}, 2290 (1986).

\bibitem{fernandez1}
D. J. Fernandez C., Int. J. Theor. Phys. {\bf 33}, 2037 (1994).

\bibitem{delgado3}
F. Delgado and B. Mielnik, J. Phys. A {\bf 31}, 309 (1997).

\bibitem{delgado4}
F. Delgado and B. Mielnik, Phys. Lett. A {\bf 249}, 359 (1998).

\bibitem{sak1}
J. J. Sakurai and J. Napolitano, {\it Modern Quantum Mechanics}. (Addison Wesley, 2010).

\bibitem{morr1}
M. A. Morrison and G. A. Parker, Aust. J. Phys. {\bf 40}, 465 (1987).

\bibitem{weigert1}
S. Weigert, J. Phys. A: Math. Gen. {\bf 30}, 8739 (1997).

\bibitem{cornwell1}
J. F. Cornwell. {\it Group Theory in Physics: an introduction} (v3.13), 2013. San Diego, California, USA: Academic Press, 1997.

\bibitem{robart1}
T. Robart, J. Leslie \& A. Banyaga. {\it Infinite Dimensional Lie Groups in Geometry and Representation Theory}. River Edge, N.J.: World Scientific, 2002. 

\bibitem{john1}
M. W. Johnson,	 M. H. S. Amin,	 S. Gildert,	 T. Lanting,	 F. Hamze,	 N. Dickson,	 R. Harris,	 A. J. Berkley,	 J. Johansson,	 P. Bunyk,	 E. M. Chapple,	 C. Enderud,	 J. P. Hilton,	 K. Karimi,	 E. Ladizinsky,	 N. Ladizinsky,	 T. Oh,	
 I. Perminov,	 C. Rich,	 M. C. Thom,	 E. Tolkacheva,	 C. J. S. Truncik,	 S. Uchaikin,	 J. Wang,	 B. Wilson and G. Rose. Nature  {\bf 473}, 194 (2011).

\bibitem{klo1}
Ch. Kloeffel and D. Loss. Annu. Rev. Condens. Matter Phys. {\bf 4}, 51 (2013).

\bibitem{loss1}
D. Loss D, D. P. DiVincenzo.  Phys. Rev. A {\bf 57},120 (1998).

\bibitem{ima1}
A. Imamoglu, D. D. Awschalom, G. Burkard, D. P. DiVincenzo, D. Loss et al. Phys. Rev. Lett. {\bf 83}, 4204 (1999).

\bibitem{kado1}
T. Kadowaki and H. Nishimori. Phys. Rev. E {\bf 58}, 5355 (1998).

\bibitem{briegel1}
H. J. Briegel, D. E. Browne, W. D\"ur, R. Raussendorf and M. Van den Nest. Quantum Information Processing, {\bf 3}, 1–5, (2004).

\bibitem{ying1}
Y. J. Zhang,  Y. J. Xia,  Z. X. Man and  G. C. Guo. Science in China Series G: Physics, Mechanics and Astronomy 
{\bf 52} (5), 700 (2009).

\bibitem{mich1}
X. Michalet1, F. F. Pinaud1, L. A. Bentolila, J. M. Tsay, S. Doose, J. J. Li, G. Sundaresan, A. M. Wu, S. S. Gambhir and S. Weiss. Science {\bf 28} 307, 538 (2005). 

\end{thebibliography}
\end{document}